\documentclass[sigconf]{acmart}

\usepackage{textcomp}                  
\usepackage{array}  
\usepackage{booktabs}  
\usepackage{ragged2e}  

\usepackage{graphicx}
\usepackage{multicol}
\usepackage{multirow}
\usepackage{xspace}
\usepackage{soul}
\usepackage{xcolor}
\usepackage{seqsplit}
\usepackage{hyperref}
\usepackage{adjustbox}
\usepackage{float}
\usepackage{url}
\usepackage{listings}
\usepackage[htt]{hyphenat}

\usepackage[scaled]{beramono}
\usepackage[T1]{fontenc}
\edef\oldtt{\ttdefault}
\usepackage[scaled]{beramono}
\usepackage[T1]{fontenc}
\renewcommand{\ttdefault}{\oldtt}

\newcommand{\scaf}[3][RGB]{%
  \begingroup
  \definecolor{scafcolor}{#1}{#2}\sethlcolor{scafcolor}%
  \hl{($\mathcal{S}$) #3}%
  \endgroup
}

\newcommand{\keyq}[3][RGB]{%
  \begingroup
  \definecolor{keyqcolor}{#1}{#2}\sethlcolor{keyqcolor}%
  \hl{($\mathcal{K}$) #3}%
  \endgroup
}

\AtBeginDocument{%
  }

\copyrightyear{2024}
\acmYear{2024}
\setcopyright{acmlicensed}\acmConference[CHI '24]{Proceedings of the CHI
Conference on Human Factors in Computing Systems}{May 11--16,
2024}{Honolulu, HI, USA}
\acmBooktitle{Proceedings of the CHI Conference on Human Factors in
Computing Systems (CHI '24), May 11--16, 2024, Honolulu, HI, USA}
\acmDOI{10.1145/3613904.3642943}
\acmISBN{979-8-4007-0330-0/24/05}

\begin{document}

\title[NL Dataset Generation Framework for Visualizations Powered by LLMs]{Natural Language Dataset Generation Framework for Visualizations Powered by Large Language Models}

\author{Hyung-Kwon Ko}
\email{hyungkwonko@gmail.com}
\affiliation{
    \institution{KAIST}
    \country{Republic of Korea}
}

\author{Hyeon Jeon}
\email{hj@hcil.snu.ac.kr}
\affiliation{
    \institution{Seoul National University}
    \country{Republic of Korea}
}

\author{Gwanmo Park}
\email{gmpark@hcil.snu.ac.kr}
\affiliation{
    \institution{Seoul National University}
    \country{Republic of Korea}
}

\author{Dae Hyun Kim}
\email{dhkim16@cs.stanford.edu}
\affiliation{
    \institution{KAIST}
    \country{Republic of Korea}
}

\author{Nam Wook Kim}
\email{nam.wook.kim@bc.edu}
\affiliation{
    \institution{Boston College}
    \country{USA}
}

\author{Juho Kim}
\email{juhokim@kaist.ac.kr}
\affiliation{
    \institution{KAIST}
    \country{Republic of Korea}
}

\author{Jinwook Seo}
\authornote{corresponding author}
\email{jseo@snu.ac.kr}
\affiliation{
    \institution{Seoul National University}
    \country{Republic of Korea}
}

\renewcommand{\shortauthors}{Ko et al.}

\begin{abstract}
  We introduce VL2NL, a Large Language Model (LLM) framework that generates rich and diverse NL datasets using Vega-Lite specifications as input, thereby streamlining the development of Natural Language Interfaces (NLIs) for data visualization. To synthesize relevant chart semantics accurately and enhance syntactic diversity in each NL dataset, we leverage 1) a guided discovery incorporated into prompting so that LLMs can steer themselves to create faithful NL datasets in a self-directed manner; 2) a score-based paraphrasing to augment NL syntax along with four language axes. We also present a new collection of 1,981 real-world Vega-Lite specifications that have increased diversity and complexity than existing chart collections. When tested on our chart collection, VL2NL extracted chart semantics and generated L1/L2 captions with 89.4\% and 76.0\% accuracy, respectively. It also demonstrated generating and paraphrasing utterances and questions with greater diversity compared to the benchmarks.
  Last, we discuss how our NL datasets and framework can be utilized in real-world scenarios.
  The codes and chart collection are available at \textcolor{blue}{\url{https://github.com/hyungkwonko/chart-llm}}.
\end{abstract}


%


\maketitle

\section{Introduction}

Recent advancements in Natural Language Processing (NLP) techniques empowered individuals with limited data analysis and visualization expertise to engage in text-based interaction and execute data visualization tasks \cite{sultanum2023datatales, shen2023data}. 
Many studies have incorporated Natural Language Interfaces (NLIs) into their systems to augment more natural and user-friendly interactions \cite{setlur2016eviza}. For example, Voder~\cite{srinivasan2018augmenting} enables querying key data insights within charts using NL sentences, significantly decreasing the reliance on manual programming for data retrieval. Furthermore, users can provide text to receive automatic recommendations for the most appropriate chart types \cite{narechania2020nl4dv, dibia2023lida}, rather than selecting effective representations manually based on graphical language criteria.

While the presence of suitable datasets modeling human behaviors is crucial in developing effective NLIs or tools for visualizations, prior work has repeatedly pointed to the scarcity of sizable pairs of high-quality datasets (chart, NL)~\cite{10.1111:cgf.14855, shen2022towards, davila2020chart, 10.2312:evp.20231072, luo2021synthesizing, srinivasan2021collecting}.
In detail, existing chart collections are occasionally synthetic~\cite{zhao2020chartseer, luo2021synthesizing}, limited in diversity (e.g., chart type)~\cite{mahinpei2022linecap, 10.1111:cgf.14855}, or are limited to simpler charts (e.g., basic bar charts, univariate line charts)~\cite{dibia2019data2vis}.
Making things worse, only a fraction of these collections (17 out of 56) is publicly accessible \cite{10.1111:cgf.14855}.
Furthermore, prior work builds the NL datasets that goes with the visualizations through crowdsourcing~\cite{srinivasan2021collecting}.
However, the process can be costly and time-consuming as it requires recruiting specific sets of target users of the system, some of whom must meet notably stringent qualification criteria.
Moreover, it is challenging to capture the language variations that arise from a diverse spectrum of user expertise, usage scenarios, and personal preferences, although this is essential for addressing the syntactic variations among the target users of the systems in the real-world \cite{gao2015datatone, setlur2016eviza, yu2019flowsense}.
What exacerbates the situation is there are multiple types of NL tasks (e.g., captioning, chart generation \& modification, and chart question-answering) where each one necessitates a new dataset tailored to the specific task or transfer knowledge.

We present a new collection of 1,981 Vega-Lite specifications (\autoref{fig:level}). This is the largest set of human-generated charts obtained from GitHub to date. It covers varying levels of complexity from a simple line chart without any interaction (i.e., simple charts) to a chart with four plots where data points are linked with selection interactions (i.e., extra complex charts) (see the charts highlighted with red stroke in \autoref{fig:level}). As we focus on amassing a richer set of charts in terms of complexity, more than 86\% of them are in complex and extra complex levels. Compared to the benchmarks, our dataset shows the highest average pairwise edit distance between specifications, which proves that the charts are highly diverse from one another. Moreover, it contains the largest number of charts with composite views, interactions (e.g., tooltips, panning \& zooming, and linking), and diverse chart types (e.g., map, grid \& matrix, diagram, etc.) (\autoref{tab:data}).

We also introduce VL2NL, a 3-stage NL generation framework that can be generalized to various NL tasks on visualizations (\autoref{fig:framework}).
First, the framework preprocesses the underlying datasets and minifies Vega-Lite specification for efficient and effective usage by an LLM.
Next, the framework leverages guided discovery~\cite{brown1994guided} so that LLMs can steer themselves to create varying NL datasets in a self-directed manner. Here, it analyzes and integrates chart semantics (e.g., mark, encoding) with our scaffolding in accordance with the characteristics of each NL dataset. Also, by answering on key questions, it autonomously concentrates on the chart's key features or propose high-level decisions.
Finally, the framework applies a score-based paraphrasing (\autoref{tab:axes}) with an LLM to simulate and include syntactic variations of human language in NL datasets.

To test VL2NL, we generated L1 captions that simply describe how the chart encodes data, L2 captions that describe the statistical properties of the data in a chart~\cite{lundgard2021accessible}, utterances for chart generation \cite{srinivasan2021collecting}, and questions for chart question answering \cite{kim2020answering, hoque2022chart}.
Our experiments showed that the accuracy of the analyzed chart semantics and generated L1/L2 captions is 89.4\% and 76.0\%, respectively.
Moreover, the generated and paraphrased NL datasets showed greater syntactic diversity in terms of 4.75 out of 6 within-distribution metrics on average.
Last, we demonstrate the application of our NL datasets in finetuning experiments, and the use of VL2NL in both fully-automatic and mixed-initiative modes within an interactive system for real-world scenarios.

The main contributions of our work are summarized as follows:
\begin{itemize}
  \item We collect 1,981 real-world Vega-Lite specifications that are diverse and go beyond simple charts;
  \item We present 3-stage NL dataset generation framework for visualizations powered by LLMs that employs guided discovery and score-based paraphrasing;
  \item We perform quantitative and qualitative analysis on the NL datasets generated by our framework.
\end{itemize}
\section{Background and Related Work}
\label{sec2}

\begin{figure*}[t]
  \centering
  \includegraphics[width=\linewidth]{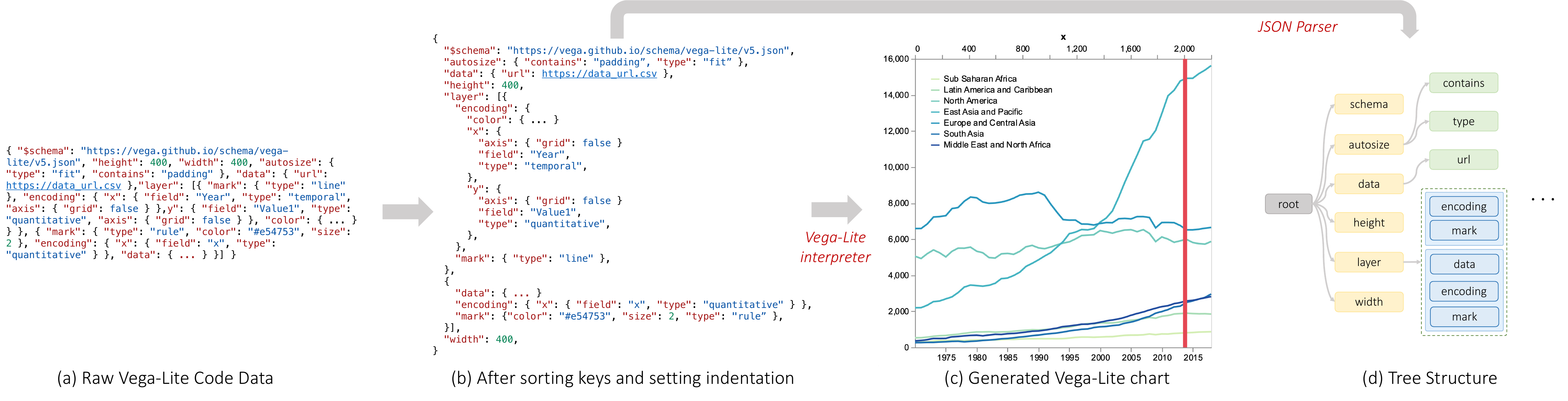}
  \caption{\textbf{Example of Vega-Lite Specification.} As previously noted in several works \cite{zhao2020chartseer, luo2021synthesizing}, Vega-Lite specification can be regarded to follow a tree structure, with its keys (i.e., properties) connected in a nested structure.}
  \label{fig:ex}
\end{figure*}

In this section, we explain Vega-Lite specification and existing chart collections. Next, we present the types of NL datasets that are of particular interest in the context of this work. Last, we explain the use of LLMs in synthesizing NL datasets.

\subsection{Chart Datasets}

According to Chen et al.'s recent survey \cite{10.1111:cgf.14855}, chart datasets are typically collected in three formats: bitmap graphics (e.g., \texttt{.png}), vector graphics (e.g., \texttt{.svg}), and programs (e.g., Vega-Lite specifications~\cite{2017-vega-lite}). Among the surveyed datasets, the majority (48 out of 56) consisted of bitmap graphics, followed by vector graphics (10 out of 56), while programs were less prevalent, comprising only five instances (some works included multiple formats).

Among many program formats, we are especially interested in Vega-Lite (\autoref{fig:ex}), which is an abstract specification that enables the creation of interactive visualizations using a high-level grammar. It is represented as a nested JSON object, consisting of numerous key-value pairs, which can be also seen as a tree structure  \cite{zhao2020chartseer, luo2021synthesizing}. Each key defined in the specification is referred to as a property \cite{wongsuphasawat2016towards}, serving a distinct role in generating charts. For example, \texttt{mark} property is used to map data to graphical elements (e.g., points, lines).

Vega-Lite provides additional advantages beyond those offered by SVG formats, since it is easy to modify and reuse for creating diverse chart variations \cite{harper2017converting}. It provides interactive features like zooming, panning, and brushing, as well as concatenating or faceting multiple plots/views. Furthermore, it support data-driven manipulation, allowing users to dynamically update the data and reflect changes in real time. It can be seamlessly converted to other formats like bitmaps and SVG \cite{vegaliteeditor}, while converting from those formats to program specifications typically requires manual effort or complex external algorithms \cite{poco2017reverse}.

There are two types of Vega-Lite benchmarks: synthetic and real-world datasets. A critical limitation of synthetic datasets lies is their reliance on pre-defined templates and rules, which leads to a high degree of repetition and a limited range of chart types and functionalities (see \autoref{tab:summary}). On the other hand, the real-world dataset reveals significant variation from one spec to another, ensuring a high level of diversity in realistic scenarios. However, they are generally much smaller in size compared to synthetic datasets \cite{10.1111:cgf.14855}.

We found three synthetic Vega-Lite datasets. In detail, Poco et al. generated 4,318 Vega specifications \cite{satyanarayan2015reactive} using the Compass recommendation engine \cite{wongsuphasawat2016towards}. They randomly selected values for a few variables (e.g., fonts, font size, legend positions, etc.) from a curated set of options. These specifications were later converted to Vega-Lite specifications in Data2Vis \cite{dibia2019data2vis}. Zhao et al. \cite{zhao2020chartseer} followed a similar approach to generate the Chartseer dataset, consisting of 9,925 specifications based on Data2Vis, although it is specifically designed for training a deep learning model and may not readily render into charts, making it less suitable for broader research adoption. The nvBench dataset \cite{luo2021synthesizing} presented 7,274 specifications, representing SQL queries as tree structures and mapping them into Vega-Lite specifications.

There are two real-world datasets that consist of human-generated specifications. For instance, Kim et al.~\cite{kim2020answering} curated 52 charts from various web sources, encompassing two chart types (bar chart and line chart). Additionally, the Vega-Lite gallery example dataset \cite{vegalitegallery}, the largest publicly available human-generated collection of Vega-Lite data, provides 716 high-quality examples with diverse chart types and interactions. However, due to the challenges associated with data collection, these datasets have a limited quantity of specifications compared to synthesized datasets. As a result, researchers often face difficulties in finding a comprehensive set of specifications for their own research purposes.

\subsection{NLIs for Data Visualization}
NLIs for data visualization have garnered significant attention due to their user-friendly nature \cite{shen2022towards, wang2022towards, srinivasan2017natural}. These interfaces allow users to focus on their tasks rather than learning how to interact with systems \cite{cox2001multi}. A recent survey paper~\cite{shen2022towards} suggested six high-level topics (e.g., visualization recommendation) to cluster tasks. They also presented a pipeline with seven stages by extending the classical information visualization pipeline \cite{card1999readings}.

To address diverse NLI tasks, we considered three types of NL datasets: captions, utterances, and questions. This choice was made based on the analysis of each topic, the number of representative works, and the relevance of NL datasets to their respective tasks.

The first NL dataset is chart caption. The captions can help people communicate and grasp insights in the charts easily, also improving the accessibility for readers of the blind and low vision people \cite{lundgard2021accessible}. A lot of research delved into this problem leveraging from templates \cite{morash2015guiding} to deep learning models \cite{obeid2020chart, qian2021generating, spreafico2020neural}.

Lundgard and Satyanarayan \cite{lundgard2021accessible} proposed a four-level classification of captions where each level contains different semantic content of the same chart: L1 provides elemental and encoded attributes, including chart type and encoding channel; L2 encompasses statistical and relational attributes such as descriptive statistics and correlation; L3 addresses perceptual and cognitive attributes, covering complex trends and patterns; L4 contains contextual and domain-specific knowledge. Recently, VisText \cite{2023-vistext} generated L2/L3 captions by training ByT5
transformer model~\cite{xue2022byt5} with crowdsourced dataset. Our work shares similarities with VisText in generating captions with varying levels. However, it differs in that we do not rely on crowdsourcing NL datasets or training machine learning models. Instead, our approach solely depends on Vega-Lite specification input and vanilla LLMs. It is worth noting that previous studies in caption generation have predominantly focused on basic chart types, as highlighted in \cite{shen2022towards}. In contrast, our work offers a generalizable solution capable of generating captions for complex and diverse charts.

The second NL dataset is utterance for chart generation. For many decades, automatically representing graphical information has been one of the important topics in information visualization \cite{mackinlay1986automating}. Many NLIs were introduced and adopted to solve multiple stages that are entangled one another for the automatic representation. The most relevant stages are 1) utterance interpretation \cite{yu2019flowsense, narechania2020nl4dv, kincaid2017nicky, hoque2017applying, srinivasan2017orko, fu2020quda, liu2021advisor, luo2021natural} and 2) mapping utterances to visual elements \cite{wongsuphasawat2015voyager, wongsuphasawat2017voyager, hu2018dive, mackinlay2007show, stolte2002polaris, kandel2012profiler, luo2018deepeye, moritz2018formalizing, vartak2015seedb}, and 3) human interaction for clarifying ambiguity or suggesting commands \cite{gao2015datatone, setlur2016eviza, narechania2020nl4dv, hoque2017applying, narechania2021diy}.

Srinivasan et al. \cite{srinivasan2021collecting} analyzed the characteristics and semantics of NL utterances employed in chart generation. According to their research, NL utterances can be classified into three types based on their structures: commands, which are instructions or systematic requests; queries, which are concise lists of keywords similar to web search queries; and questions, which are data-driven inquiries that users wish to visualize. In our work, we generate all three types of utterances, incorporating heightened syntactic diversity for a comprehensive evaluation.

The last NL dataset is question. Chart Question Answering is a popular task in both machine learning \cite{liu2022deplot, liu2022matcha, masry2022chartqa, chaudhry2020leaf} and human-computer interaction \cite{kim2020answering} communities. This popularity stems from its effectiveness in eliciting insights and aiding in decision-making processes \cite{hoque2022chart}.

Kim et al.~\cite{kim2020answering} investigated the semantics used in the questions by collecting 629 crowd-sourced questions and provided two orthogonal dimensions. First axis is lookup or compositional, which is whether to retrieve a single value or using multiple mathematical operations. Second axis is visual or non-visual, which is whether to reference visual chart features or not. These question types are all focusing on retrieving factual short answers. In our work, we target five different types of questions, including the aforementioned types as well as the open-ended question type \cite{hoque2022chart}, which encourages deeper reflection on the underlying reasons or causes behind specific events or patterns.

\subsection{LLMs and NL Datasets}
Many past research typically have used crowdsourcing to collect varying types of NL datasets (e.g., captions, utterances, questions, etc.) by asking crowd workers to come up with generation queries using available chart datasets \cite{srinivasan2021collecting, luo2021synthesizing, kim2020answering}.
However, this approach is frequently time-consuming and costly \cite{wu2023webui, deka2017rico}, which can adversely affect the scalability of datasets. It is prone to issues such as participant laziness and the collection of subpar queries \cite{bernstein2010soylent}.
To ensure a consistent performance among workers, it is essential to simplify the tasks and making them easy to follow, thereby preventing workers from feeling overwhelmed or fatigued during the study, as recommended by Kittur et al.~\cite{kittur2011crowdforge}.
With all these efforts, such crowdsourced NL datasets are often fragmented, posing challenges for researchers seeking to apply them to their own tasks.
The characteristics of NL queries designed for each task can vary significantly, making a single NL dataset unsuitable for other tasks.
This motivates the need for a unified and adaptable framework that can generate NL datasets tailored to any specific NLIs for data visualization research.

As LLMs are known to simulate human behavior \cite{park2023generative} and have become more prevalent due to their powerful performance, researchers are increasingly using generated NL datasets to train smaller-sized language models for specific tasks \cite{schick2021generating, meng2022generating, ye2022zerogen}.
This training strategy is known as `teaching via data'~\cite{li2023flexkbqa}. Here, LLMs, acting as teacher models, generate synthetic datasets which are then used to train smaller-sized models, referred to as students, designed for specific tasks.
This method is adopted to increase the performance of different tasks like knowledge-based question answering~\cite{li2023flexkbqa}, symbolic language generation (e.g., SQL query)~\cite{ye2023generating}, and semantic parsing~\cite{rosenbaum2022clasp}.
Our work aligns with this trend, aiming to assist researchers in developing NLIs for data visualization by generating the necessary NL datasets using LLMs.

\section{Vega-Lite Dataset}
\label{sec4}

\begin{figure*}[t]
  \centering
  \includegraphics[height=\textwidth]{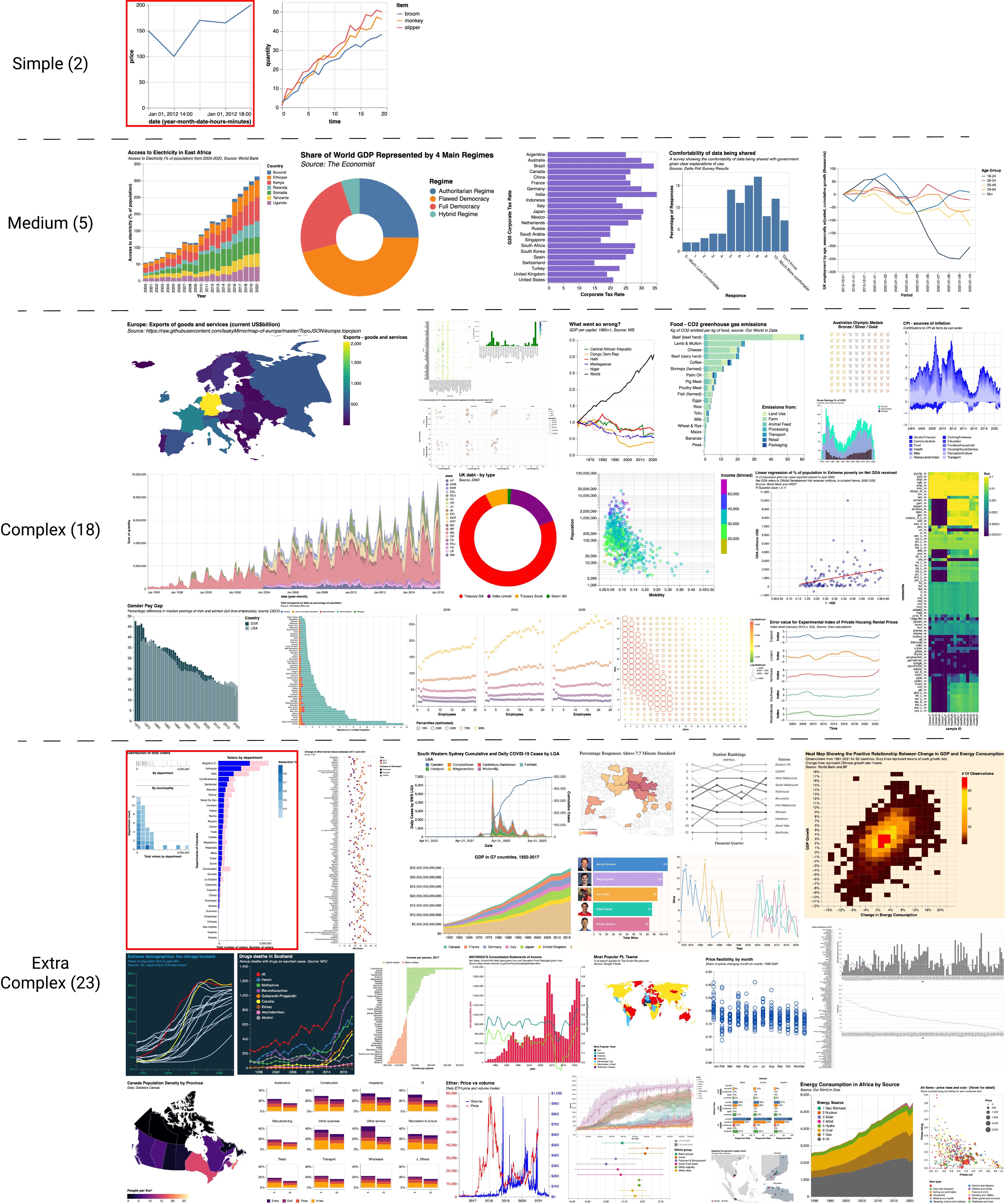}
  \caption{\textbf{Vega-Lite dataset divided by their complexity levels: simple, medium, complex, extra complex.}
  These 48 charts were selected via stratified sampling and used in our evaluation (\autoref{sec5}).
  The level is divided based on the number of keys each specification contains.
  The number of keys, which are the criteria for dividing the levels, are set based on the quartiles (Q1, Q2, Q3) of Vega-Lite example gallery dataset~\cite{vegalitegallery}.
  }
  \label{fig:level}
\end{figure*}

\begin{table}[t]
\caption{\textbf{Summary of the Vega-Lite dataset construction process.} First we collect all possible cases of URLs including Vega-Lite specifications (a). Next, we have filtered unique URLs that are allowed to redistribute for academic purpose (b, c). Finally we iteratively inspect each specification manually to check whether it is valid and unique, since we want to collect charts with a high level of diversity (d).}
\label{tab:summary}
\begin{center}
\begin{tabular}{lr}
\toprule
 & \# of URLs / Vega-Lite specs\\ 
\midrule
(a) URLs crawled & 67,789\\ 
(b) URLs w/o duplicate & 18,420\\ 
(c) URLs w/ license & 7,408\\ 
(d) Specs after manual inspection & 1,981\\ 
\bottomrule
\end{tabular}
\end{center}
\end{table}

\begin{table*}[t]
\caption{\textbf{Summary statistics of our dataset and benchmark datasets that are publicly available.} Two types of datasets are presented: synthetic and real-world datasets. The best statistics within each type are highlighted in bold, while the best statistics across all datasets are also underscored.}
\label{tab:data}
\begin{center}
\resizebox{\linewidth}{!}{%
\begin{tabular}{llrrr|rrr}
\toprule
\multirow{2.5}{*}{Type} & \multirow{2.5}{*}{Evaluation Metric / Criteria} & \multicolumn{3}{c}{\textbf{Synthetic data (machine-generated)}} & \multicolumn{3}{c}{\textbf{Real-world data (human-generated)}} \\
\cmidrule{3-8}
& & Data2Vis~\cite{dibia2019data2vis} & Chartseer~\cite{zhao2020chartseer} & nvBench~\cite{luo2021synthesizing} & Kim et al.~\cite{kim2020answering} & Gallery~\cite{vegalitegallery} & Ours \\
\midrule
Quantity & \# of specs & 4,318 & \textbf{\underline{9,897}} & 6,680 & 52 & 709 & \textbf{1,981} \\
\midrule
\multirow{9}{*}{Complexity} & Total \# of keys across specs & 101,881 & \textbf{\underline{147,676}} & 98,074 & 769 & 26,469 & \textbf{107,802} \\
& Average \# of keys in a spec & \textbf{24} & 15 & 15 & 15 & 37 & \textbf{\underline{54}} \\
\cmidrule{2-8}
& Simple (key $\leq$ 16) & 0 & 6,164 & \textbf{\underline{6,354}} & 41 & \textbf{186} & 73 \\
& Medium (key $\leq$ 24) & \textbf{\underline{4,318}} & 3,733 & 326 & 10 & 170 & \textbf{199} \\
& Complex (key $\leq$ 41) & 0 & 0 & 0 & 1 & 179 & \textbf{\underline{733}} \\
& Extra complex (key $>$ 41) & 0 & 0 & 0 & 0 & 174 & \textbf{\underline{976}} \\
\cmidrule{2-8}
& Average depth of JSON & \textbf{4.00} & 3.00 & 3.48 & 3.13 & 5.01 & \textbf{\underline{5.19}} \\
& Average branching factor & 1.22 & \textbf{\underline{1.44}} & 1.18 & 1.17 & \textbf{1.41} & 1.38 \\
\midrule
\multirow{5.5}{*}{Diversity} & Total \# of unique keys & \textbf{24} & 12 & 18 & 31 & 275 & \textbf{\underline{362}} \\
& Average pairwise edit distance & \textbf{122.62} & 75.90 & 48.18 & 129.51 & 1,096.11 & \textbf{\underline{1,549.48}} \\
\cmidrule{2-8}
& Composite views & 0 & 0 & 0 & 0 & 136 & \textbf{\underline{746}} \\
& Interaction (e.g., zoom, pan) & 0 & 0 & 0 & 0 & 188 & \textbf{\underline{1,010}} \\
& \# of chart types & \textbf{6} & \textbf{6} & 4 & 2 & \textbf{\underline{10}} & \textbf{\underline{10}} \\
\bottomrule
\end{tabular}
}
\end{center}
\end{table*}


We have collected a new set of 1,981 real-world Vega-Lite specifications.
In this section, we present the details of our data collection process.

\subsection{Dataset Construction}

\subsubsection{Search Queries.}
We utilize the GitHub API\footnote{https://docs.github.com/en/rest} to create our Vega-Lite dataset. Due to the API's limitation of providing up to 1,000 results per search query, we employ various techniques, as we elaborate below, to crawl Vega-Lite specifications in a mutually exclusive and exhaustive manner to the best of our abilities.

When building search queries, we use the keyword \path{https://vega.github.io/schema/vega-lite/[version]} to indicate the version of the specification that Vega-Lite uses for rendering purposes. We collect versions from v2 to v5: there are no v1 data to be found. To partition the query into a more fine-grained manner, we use keywords such as \texttt{.csv} and \texttt{.json} to gather specifications with external links. Similarly, we employ keywords like \texttt{values} and \texttt{datasets} to identify ones with internally embedded data. We also leverage additional keywords using the main properties defined in the version 5 Vega-Lite specification\footnote{https://github.com/vega/schema}. These properties encompass essential elements for creating a single plot, including \texttt{data}, \texttt{transform}, \texttt{mark}, and \texttt{encoding}, while there are properties like \texttt{layer}, \texttt{facet}, \texttt{concat}, and \texttt{repeat}, which are specifically relevant to visualizing \textit{composite} views \cite{2017-vega-lite} (e.g., layered plots, trellis plots, or multiple views). A comprehensive list of the properties we use can be found on the official documentation page\footnote{https://vega.github.io/vega-lite/docs}.


\subsubsection{Inclusion and Exclusion Criteria.}
We target files with extension \texttt{.json}, \texttt{vg.json}, \texttt{.vl.json}, \texttt{.vl}, and \texttt{.vg} which denotes Vega-Lite specifications. We also examine HTML and JavaScript files containing Vega-Lite specifications manually to get additional specifications. Throughout the process, we exclude forked repositories to prevent redundancy. We also filter out any data from the benchmark datasets, such as Vega-Lite gallery~\cite{vegalitegallery}.

\subsubsection{Post-processing.}
To obtain a large number of unique sets of Vega-Lite specifications, we follow a step-by-step approach.
During the initial stage, a total of 67,789 URLs are collected. Despite efforts to ensure a mutually exclusive and comprehensive set of specifications, duplicate URLs are identified and removed, resulting in 18,420 unique URLs. Each URL is scrutinized to verify the license of the corresponding repository, ensuring compliance with copyright regulations for academic redistribution. This process yields 7,408 URLs. Lastly, we verify their validity using the Vega-Lite editor~\cite{vegaliteeditor}. This involves identifying the URLs of the datasets used by each specification and making necessary modifications, ranging from minor adjustments such as closing unclosed brackets to more significant ones like debugging the entire code, in order to achieve successful rendering. 
An overview of the post-processing and the number of URLs and specifications obtained at each stage can be found in \autoref{tab:summary}. 
Our chart collection is publicly accessible via the following link: \textcolor{blue}{\url{https://hyungkwonko.info/chart-llm-data}}. 

\subsection{Quantitative Analysis}

\subsubsection{Benchmarks.} 
We compare three synthetic and two real-world Vega-Lite datasets \cite{vegalitegallery, dibia2019data2vis, kim2020answering, zhao2020chartseer, luo2021synthesizing} described in \autoref{sec2}. To ensure a fair comparison, we implement a process to remove exact code duplication within each benchmark. In detail, each specification is sorted in alphabetical order by the keys and edited to maintain consistent indentation. Next, we convert each file into a hash where files with identical hashes are subsequently removed from the dataset. Following this procedure, the number of specifications in Chartseer dataset decrease from 9,917 to 9,897, nvBench decrease from 7,241 to 6,680, and the Vega-Lite gallery example dataset decrease from 716 to 709.

\subsubsection{Quality Metrics.}
To comprehensively assess the Vega-Lite datasets, we consider three different aspects: quantity, complexity, and diversity. Initially, we count the number of collected specifications to determine the overall quantity of Vega-Lite specifications, as previously done by Luo et al.~\cite{luo2021synthesizing}. However, we argue that additional metrics are necessary to gauge the quality of the Vega-Lite dataset. This is because some specifications include only mandatory properties to construct a single plot without any interaction (e.g., \texttt{data}, \texttt{encoding}, \texttt{mark} for a simple bar chart), while others contain multiple plots or views linked by varying interactions. Therefore, the number of keys in a specification can highly differ depending on whether it includes properties for data pre-processing (e.g., \texttt{aggregate}, \texttt{calculate}, etc.), interactivity (e.g., \texttt{bind}, \texttt{select}, etc.), or composite views (e.g., \texttt{concat}, \texttt{repeat}, etc.). We can expect the Vega-Lite specification becomes more complex as the number of defined properties increases. Therefore, we propose a new standard to understand the complexity of a Vega-Lite dataset by counting the total number of keys present across all specifications and the average number of keys in a singe specification. To ensure a fair comparison, we only consider keys defined in the version 5 specification. We also ignore keys associated with internally embedded datasets, such as \texttt{values} and \texttt{datasets,} along with their corresponding keys. In addition to this, we also measure the average depth and branching factor of the JSON structure as they are commonly adopted to evaluate the complexity of a JSON file.

We found no metrics to quantify the diversity of chart dataset \cite{10.1111:cgf.14855}. Therefore, we also propose metrics for gaining insights into the diversity of dataset in terms of both the range of properties within the entire dataset and the variance between individual specifications. 
Specifically, we count the number of unique keys employed across the entire dataset and calculate the average pairwise edit distance among all possible pairs of specifications. 
The number of unique keys indicates how many distinct properties that can be defined in a Vega-Lite specification are used across the specifications. For example, if a handful of unique keys are used within the dataset, this indicates a restricted recurrence of only a few properties. In turn, it likely signifies a low level of diversity. The average pairwise edit distance provides an overview of the dissimilarity between each pair at the code level. To perform this analysis, we sort the keys alphabetically, replace their corresponding values with empty values, and exclude keys associated with embedded datasets, as mentioned earlier. 

\subsubsection{Complexity Levels.}
We observe that the existing criteria used to establish the complexity levels of charts are somewhat subjective and may not possess broad applicability \cite{luo2021synthesizing, kantharaj2022chart, lundgard2021accessible}.
Instead, we suggest using the number of keys as a criterion for categorizing the complexity levels of charts, particularly in the context of Vega-Lite specifications. This is because, as explained above, the number of properties increases proportionately to the number of keys in a specification. To establish the standard number of keys, we refer to the Vega-Lite example gallery dataset \cite{vegalitegallery} and calculate the quartiles (Q1, Q2, Q3) based on the distribution of the number of keys. These quartiles, specifically 16, 24, and 41, are utilized as reference points to divide the specifications' level of complexity. For instance, a specification with a total number of keys less than or equal to 16 is classified as `simple' complexity. Likewise, a specification with a total number of keys greater than 16 and less than or equal to 24 is classified as `medium' complexity (\autoref{fig:level}).

\subsubsection{Composite View, Interactivity, and Chart Type Distribution.}
We choose three additional factors by referring to previous works \cite{li2022diverse, borkin2013makes} to further assess the quality of the datasets. First, we examine the presence of composite views, which offer diverse perspectives on the same data simultaneously \cite{chen2020composition}. Secondly, considering the benefits of collecting Vega-Lite specifications over static bitmap images, we count the number of charts that incorporate interactive techniques such as tooltips, zooming, and brushing. Lastly, we evaluate the number of charts types based on the taxonomy proposed by Borkin et al. \cite{borkin2013makes}.

\subsubsection{Results.}
We present the results in terms of quantity, complexity, and diversity, highlighting the superiority of our dataset compared to the benchmarks. Regarding quantity, all three synthetic datasets demonstrate a higher number of specifications compared to the other three real-world datasets. Among all datasets, Chartseer shows the highest number of specifications (i.e., 9,897), while our dataset has 1,981 specifications which outnumbers the other real-world datasets in terms of quantity.

In terms of complexity, our dataset ranks the first in average number of keys in a single specification (i.e., 54) and the second in total number of keys across specifications (i.e., 107,802), which is 1.4 and 4.0 times larger than the largest previous real-world Vega-Lite dataset, respectively. Chartseer presents the highest total number of keys across specifications (i.e., 147,676) with the smallest average number of keys per specification (i.e., 15) among all datasets. Our dataset includes the highest number of specifications classified as complex (i.e., 733) and extra complex (i.e., 976), while all synthetic datasets do not contain any specifications in the complex and extra complex level. Data2Vis and nvBench demonstrate the largest number of specifications classified as medium (i.e., 4,318) and easy (i.e., 6,354), respectively. Our dataset also exhibits the highest average depth of JSON structure (i.e., 5.19), while Chartseer showcases the highest average branching factor (i.e., 1.44).

Lastly, with respect to diversity, our dataset demonstrates the largest total number of unique keys and the highest average pairwise edit distance among all datasets. Furthermore, our dataset includes the largest number of specifications featuring composite views (i.e., 1,010) and interactions (i.e., 746), exceeding the Vega-Lite gallery dataset by 1.8 and 5.3 times, respectively. None of the synthetic datasets or Kim et al.'s dataset include specifications with composite views and interactions. Both our dataset and the Vega-Lite gallery dataset cover the widest variety of chart types, encompassing ten types: Area, Bar, Circle, Diagram, Distribution, Grid \& Matrix, Line, Map, Point, and Trees \& Networks. Please refer to \autoref{tab:data} for detailed results.

\section{VL2NL: NL Generation Framework}
\label{sec3}

\begin{figure*}[t]
  \centering
  \includegraphics[width=\linewidth]{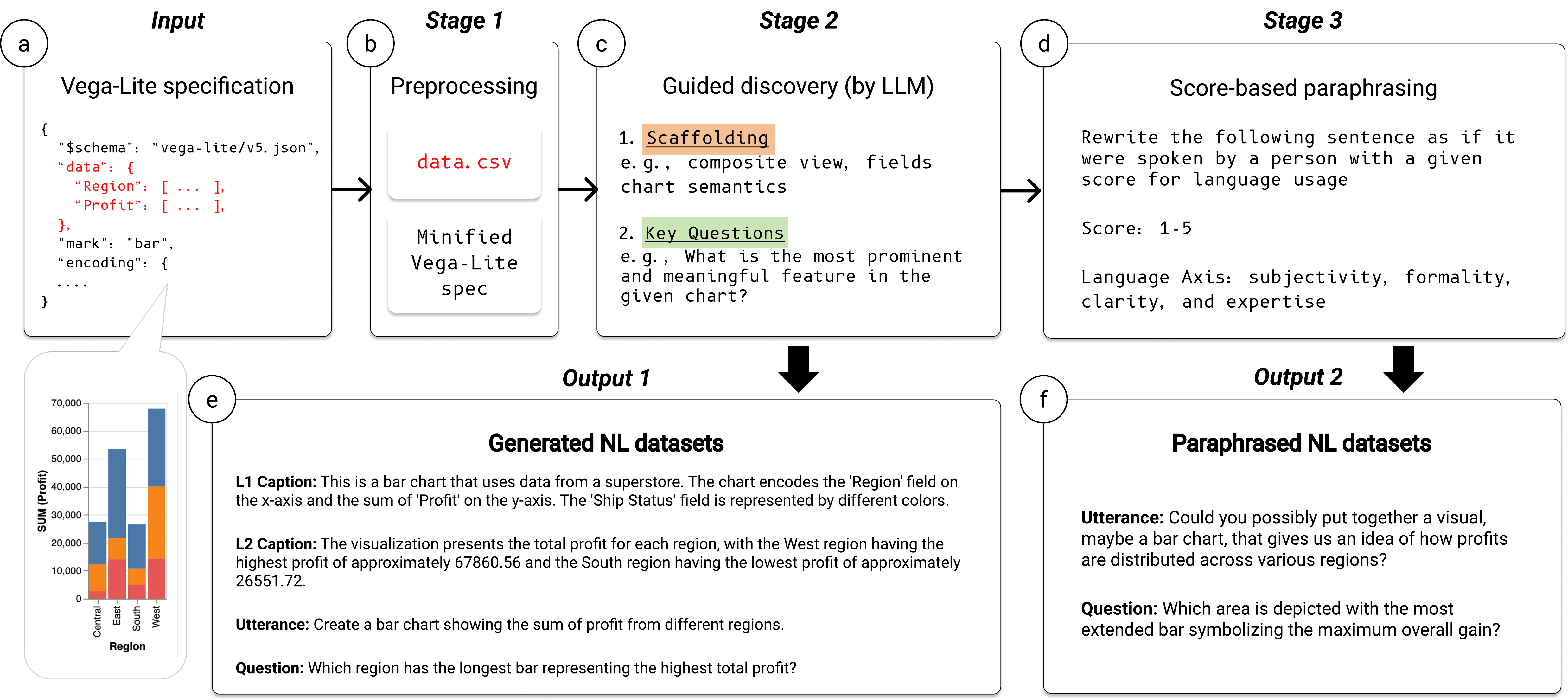}
  \caption{\textbf{LLM Framework to Generate NL Datasets for Visualizations.} We start by (b) preprocessing underlying datasets and minifying Vega-Lite specifications. Subsequently, (c) we employ scaffolding and key questions, (e) to generate NL datasets like L1/L2 captions, utterances, and questions. (d) This is followed by score-based paraphrasing, (f) allowing us to produce syntactically paraphrased NL datasets.}
  \label{fig:framework}
\end{figure*}

\begin{table}[t]
\caption{\textbf{Prompting techniques to generate each NL dataset.} Each prompt is designed by choosing the most appropriate techniques considering their different characteristics.}
\label{tab:framework}
\begin{center}
\resizebox{\linewidth}{!}{%
\begin{tabular}{llcccc}
\toprule
Target & Technique & L1 caption & L2 caption & Utterance & Question\\
\midrule
\multirow{2}{*}{Semantic} & \scaf[RGB]{254,190,140}{Scaffolding} & O & - & O & -\\
& \keyq[RGB]{182,226,161}{Key question} & - & O & O & O\\
\midrule
Syntactic & Paraphrasing & - & - & O & O\\
\bottomrule
\end{tabular}
}
\end{center}
\end{table}

\begin{table*}[t]
\caption{\textbf{Results of automatic qualitative coding~\cite{hamalainen2023evaluating}.} From previous NL datasets of captions, utterances, and questions, we identified four language axes of syntactic diversity: subjectivity, formality, clarity, and expertise. The top five most frequently occurring codes within each axis are presented along with their respective frequencies in parentheses.
}
\label{tab:coding}
\begin{center}
\begin{tabular}{c|cccc}
\toprule
Axes & Formality & Clarity & Expertise & Subjectivity\\ 
\midrule
Directions & Colloquial/Standard & Implicit/Explicit & Non-technical/Technical & Subjective/Objective\\ 
\midrule
\multirow{5}{*}{\shortstack{Example\\codes}} & interrogative form (540) & specificity (221) & economic (159) & descriptive (611)\\
& formal (384) & specific (91) & geographical context (125) & negative connotation (58)\\
& passive voice (211) & ambiguity (68) & financial (106) & subjectivity (22)\\
& analytical (195) & conciseness (64) & business-oriented (81) & negative (15)\\
& command-oriented (166) & abstract (63) & business terminology (65) & third person perspective (11)\\
\bottomrule
\end{tabular}
\end{center}
\end{table*}

\begin{table*}[t]
\scriptsize
\caption{\textbf{Example of score-based paraphrasing with two axes.} We used a sample LLM-generated utterance to create the chart shown in \autoref{fig:framework}. We perform linear interpolation to manipulate formality and expertise scores from 1 to 5, one at a time to generate 25 paraphrased sentences. While the content (i.e., semantics) of the utterance remains consistent, the tone and voice (i.e., syntax) change linearly in accordance with the provided scores.}
\label{tab:axes}
\begin{center}
\resizebox{\linewidth}{!}{%
\begin{tabular}{c|c|p{2cm}|p{2cm}|p{2cm}|p{2cm}|p{2cm}}
\specialrule{0.0pt}{0.3em}{0.3em}
\multicolumn{7}{c}{\textbf{Sample Utterance:} Create a bar chart showing the sum of profit from different regions.}\\
\specialrule{0.0pt}{0.3em}{0.3em}
\addlinespace
\toprule
& & \multicolumn{5}{c}{Expertise (Non-technical: 1, Technical: 5)}\\
\midrule
& & \multicolumn{1}{c|}{1} & \multicolumn{1}{c|}{2} & \multicolumn{1}{c|}{3} & \multicolumn{1}{c|}{4} & \multicolumn{1}{c}{5}\\
\midrule
& \multirow{5}{*}{1} & Hey, can you whip up a bar graph showing how much dough we've made from different places? & Hey, can you make a bar graph showing the total profit from different regions? & Can you put together a bar chart showing the aggregate profit from various geographical areas? & Can you construct a bar chart illustrating the cumulative profit derived from distinct regions? & Can you generate a bar chart delineating the summation of fiscal gain from disparate geographical sectors?\\
\cmidrule{2-7}
\multirow{3}{*}{\rotatebox[origin=c]{90}{Formality (Colloquial: 1, Standard: 5)}} & \multirow{5}{*}{2} & Could you create a bar chart that shows how much money we've made from different places? & Could you create a bar chart that shows the total profit from different regions? & Could you create a bar chart that illustrates the aggregate profit from various geographical areas? & Could you create a bar chart that delineates the cumulative profit derived from distinct regions? & Could you create a bar chart that represents the summation of fiscal gain from disparate geographical sectors?\\
\cmidrule{2-7}
& \multirow{5}{*}{3} & Please create a bar chart showing how much money we've made from different places. & Please create a bar chart showing the total profit from different regions. & Please create a bar chart illustrating the aggregate profit from various geographical areas. & Please create a bar chart delineating the cumulative profit derived from distinct regions. & Please create a bar chart representing the summation of fiscal gain from disparate geographical sectors.\\
\cmidrule{2-7}
& \multirow{6}{*}{4} & It is requested that you create a bar chart showing the money made from different places. & It is requested that you create a bar chart showing the total profit from different regions. & It is requested that you create a bar chart illustrating the aggregate profit from various geographical areas. & It is requested that you create a bar chart delineating the cumulative profit derived from distinct regions. & It is requested that you create a bar chart representing the summation of fiscal gain from disparate geographical sectors.\\
\cmidrule{2-7}
& \multirow{6}{*}{5} & You are required to construct a bar chart demonstrating the monetary gain from various locations. & You are required to construct a bar chart demonstrating the total profit from different regions. & You are required to construct a bar chart illustrating the aggregate profit from various geographical areas. & You are required to construct a bar chart delineating the cumulative profit derived from distinct regions. & You are required to construct a bar chart representing the summation of fiscal gain from disparate geographical sectors.\\
\bottomrule
\end{tabular}
}
\end{center}
\end{table*}

The goal of our framework is to generate high-quality NL datasets using Vega-Lite specifications and prompt engineering.
VL2NL consists of three stages (\autoref{fig:framework}). First it preprocesses underlying datasets (e.g., \texttt{csv}) and minifies the Vega-Lite specifications. Next, it identifies relevant and accurate information through guided-discovery. Last, it increases syntactic diversity using score-based paraphrasing.
To generate each type of NL dataset, we design each prompt to be maximally helpful by selecting the most appropriate strategies (\autoref{tab:framework}).

\subsection{Pre-processing Vega-Lite Specifications}
Using raw Vega-Lite specifications is not appropriate for prompting, because some of them include the dataset they use within the specification, resulting in excessively file length. Therefore, we save the data as an external files with the most suitable data formats (e.g., \texttt{.csv}, \texttt{.json}). Subsequently, the location of the saved files is overwritten with their URLs, rather than being embedded in the specification. In our current implementation of the framework, we only support \texttt{.csv} data format. Therefore, we have converted \texttt{.json} files into \texttt{.csv} files. Last, we minify the Vega-Lite specifications by removing all line breaks and indentations to reduce the number of tokens sent through API usage.

\subsection{Ensuring Accuracy and Relevance}
Our framework leverages the concept of guided discovery~\cite{brown1994guided} based on Chain-of-Thought prompting~\cite{wei2022chain} to harness the maximum reasoning capability of LLMs. 
We employ two strategies of guided discovery: providing scaffolds~\cite{hmelo2007scaffolding}. and posing key questions~\cite{de1998scientific}.
To analyze and integrate the chart semantics necessary for generating a specific NL dataset, we assist LLMs by offering scaffolds.
Additionally, we furnish LLMs with key questions to guide their self-directed progress. This maximizes the use of LLMs' reasoning abilities, allowing them to make decisions on which aspects to focus on and delve into when creating a particular data.
Below, we denote each step where we italicize the relevant phrases, and utilize symbols for \scaf[RGB]{254,190,140}{scaffolding} and \keyq[RGB]{182,226,161}{key question} to make them easily identifiable.

To demonstrate how we can ensure relevance and accuracy in generating different types of NL datasets, we have selected three datasets commonly used in NLIs for data visualization research. We made these selections based on their significance in conjunction with related tasks, as indicated in a recent survey paper~\cite{shen2022towards}: captions (L1, L2), utterance (command, query, question), question (visual-lookup, visual-compositional, nonvisual-lookup, nonvisual-compositional, open-ended). The detailed generation process for each NL dataset can differ from one another. Here, we design each step in prompting to be merged or separated when generating different NL datasets so they can best capture each of their characteristics. We only explain the high-level descriptions of each, and the detailed and full prompting used for generating each NL dataset is presented in \autoref{app:prompt}.


\subsubsection{L1 Caption}
Considering real-world Vega-Lite specifications, we first understand whether the given chart is \scaf[RGB]{254,190,140}{a composite view} (e.g., layered, trellis, and multiple views), to enable top-down analysis of each chart one by one.
The prompt follows a template with three questions to answer: Is it a composite view?; If it is, identify its type among layered, trellis, and multiple views; and determine the number of plots in the chart.
Next, it analyzes each chart individually based on the provided scaffold of \scaf[RGB]{254,190,140}{chart semantics}: Data, Transform, Mark, Chart-Type, Encoding, Style, and Interaction, using the information about composite view.
Here, we access the underlying dataset to provide \scaf[RGB]{254,190,140}{fields} that are presented in the Vega-Lite specification, along with their synonyms (i.e., \scaf[RGB]{254,190,140}{titles} also found in the same Vega-Lite specification) and their \scaf[RGB]{254,190,140}{unique values} if they are categorical variables.
After analyzing all of these semantics, the LLM finally generates the L1 caption by combining them.

\subsubsection{L2 Caption}
L2 captions, unlike L1 captions that provide an overall description of the chart, offer the flexibility to selectively focus on specific features that capture the viewer's interest. To craft informative and insightful captions, we follow a structured approach centered around a key question: \keyq[RGB]{182,226,161}{What is the most prominent and meaningful feature in the given chart?} Once we identify this feature, we delve deeper by exploring the mathematical operations required to analyze it: \keyq[RGB]{182,226,161}{What is the mathematical operation(s) required to describe the feature?} Subsequently, based on these operations, we generate \keyq[RGB]{182,226,161}{a series of questions to analyze the feature} (e.g., for the simple line chart with a red border in \autoref{fig:level}, the following questions are generated: What was the highest stock price of Google?; What was the lowest stock price of Google?; What is the difference between the highest and lowest stock prices of Google?). This process allows us to create captions that provide valuable insights into the chart's content.
When answering questions, we utilize backing datasets and LangChain~\cite{Chase_LangChain_2022} to perform required calculation that bolster the factual integrity of the generated captions.
This step is crucial, as large language models (LLMs) have been known to produce hallucinations in response to mathematical problems \cite{ji2023survey}. Once each question is answered, the collected information is subsequently incorporated into the final prompting stage for generating L2 captions. It's important to note that, unlike previous work~\cite{2023-vistext}, we do not use any of the L1 captions when generating L2 captions. Instead, this is performed as an independent process.

\subsubsection{Utterance}
Similar to L1 captions, we begin by analyzing whether the chart is a \scaf[RGB]{254,190,140}{composite view}. We then proceed to generate \scaf[RGB]{254,190,140}{instructions} for each plot independently. This process entails creating a comprehensive set of step-by-step instructions for constructing each plot. To enhance readability and user-friendliness, we ensure that each instruction focuses on a single specific action, aligning with the same semantics used when generating L1 captions. For example, in the case of the simple line chart with a red border shown in \autoref{fig:level}, the following instructions are generated:
\begin{itemize}
  \item Data: Use Google's stock price data;
  \item Chart-Type: Create a line chart;
  \item Mark: Use a line mark; 
  \item Encoding: Encode the x-axis with the date field, using a temporal type and a time unit of year, month, date, hours, and minutes, and scale it using UTC;
  \item Encoding: Encode the y-axis with the price field, using a quantitative type.
\end{itemize}
However, it is important to note that the generated instructions may sometimes feature overly technical variable names from the chart, which might not align with users' NL usage patterns.
In such cases, we leverage information from synonyms found in the underlying dataset.
Specifically, we use \scaf[RGB]{254,190,140}{title} of the Vega-Lite specification and \scaf[RGB]{254,190,140}{values} from the fields to replace the technical terms, resulting in more user-friendly instructions.

Next, we ask a key question to \keyq[RGB]{182,226,161}{identify primary and secondary information}. In this context, we anticipate that LLMs are able to automatically prioritize crucial semantics to paint a comprehensive picture of the chart, such as chart type or encoding, over additional instructions like style or interaction. This thought is based on Wang et al.'s observations~\cite{wang2022towards}, who noted that the typical workflow for creating visualizations often starts with this information (e.g., 'show me the price over time as a line chart'). Once we have all these components ready, we proceed to generate each type of utterance one by one, adhering to \scaf[RGB]{254,190,140}{specific rules} for each type. For commands, we employ the imperative voice. For queries, we use only variables, fields, attributes, mathematical formulas, abbreviations, and prepositions, while avoiding verbs and articles. For questions, we formulate inquiries in the form of questions. Across all types, we maintain the following rules: express each utterance in a single sentence, utilize only primary information, and keep the language concise and straightforward.

\subsubsection{Question}
In general, we conduct chart question answering to facilitate decision-making \cite{kim2020answering}. Thus, the process involves analyzing charts through a question-answering, which ultimately leads to a conclusion and informs the decision-making. To generate questions, we employ a reverse thought process. This entails first identifying the decisions that can be derived from the charts (i.e., \keyq[RGB]{182,226,161}{What higher-level decision can be made by analyzing this chart?}), followed by formulating a possible conclusion that leads to such a decision (i.e., \keyq[RGB]{182,226,161}{What is a possible conclusion that can be reached from this decision?}). Finally, we determine what needs to be analyzed (i.e., \keyq[RGB]{182,226,161}{What specific value can be retrieved to reach this conclusion? What are the mathematical operations to reach the conclusion?}). We generate non-visual lookup and compositional questions using the provided values and mathematical operations. To transform these into visual questions, we identify the necessary visual attributes and incorporate them into the generated questions (i.e., \keyq[RGB]{182,226,161}{What visual attributes are required to paraphrase this question?}). Finally, we formulate an open-ended question designed to lead to the same conclusion obtained in the previous step.

\subsection{Increasing Syntactic Diversity}
\label{sec3:paraphrase}

\subsubsection{Automatic Qualitative Coding}
Before increasing the syntactic diversity of NL datasets, we need to analyze which meaningful axes of diversity to address. To this end, we collected sample NL sentences from existing sources, which consist of 2,147 captions, 893 utterances, and 629 questions \cite{lundgard2021accessible, srinivasan2021collecting, kim2020answering}. Next, following the automatic coding process that Hämäläinen et al. have proposed \cite{hamalainen2023evaluating}, we utilized these sample sentences to conduct a thematic analysis using LLMs, generating five different codes for each caption, utterance, and question (see the prompt in \autoref{app:aqc}). We manually checked the generated codes to eliminate irrelevant and erroneous ones, resulting in 15,271 valid codes out of 18,345. Then, we retained 2,759 unique codes and vectorized them using Sentence-Bert~\cite{reimers-2019-sentence-bert}. Afterwards, we applied dimensionality reduction technique to project them into a lower dimensional space using UMAP~\cite{mcinnes2018umap}, reducing the 100-dimensional vectors to 5-dimensional vectors. Next, we employed HDBSCAN~\cite{mcinnes2017hdbscan} to cluster them into a few classes for detailed investigation. We aggregated clusters into a higher-level cluster, except for the codes that are not clustered through HDBSCAN, to derive the final themes.

We identified a total of six themes, but selected four meaningful axes related to NL syntax--clarity, expertise, formality, subjectivity (\autoref{tab:coding}). Two themes were removed--1) measurement, and 2) chart and data analytics--as they are not directly related to the syntax of NL datasets but rather to the semantic properties of charts. Clarity represents a language axis with two opposite meanings—implicit and explicit. Implicit language relies on context, shared knowledge, and non-verbal cues to convey meaning, while explicit language is clear and direct, leaving little room for interpretation or misunderstanding. The expertise axis also has two opposite meanings—non-technical and technical. Technical language includes specialized terminology and jargon, whereas non-technical language is more accessible to a general audience and avoids the use of complex terms. Formality, the third language axis, ranges from colloquial, which is informal and used in everyday conversation, to standard, which follows established rules and conventions. Finally, the subjectivity axis encompasses subjective language, which expresses personal opinions, feelings, or judgments, and objective language, which presents facts or information without bias or personal interpretation.

\subsubsection{Score-based Paraphrasing}

Our paraphrasing technique is inspired by a linear interpolation in the latent space for image generation and manipulation as demonstrated in many system and application papers \cite{mozaffari2022ganspiration, abdrashitov2020interactive, aoki2022emoballoon, wetoon}.
This technique enables a smooth transition from one expression to another by focusing on creating controllable and meaningful syntactic variations of a single sentence.
The key idea is that we assign language axesand employ a five-point Likert-scale to each. Here, we focus on altering only the sentence's syntax, while maintaining its meaning.
In detail, we provide LLMs with a sentence we want to paraphrase, and an explanation about one of the defined axes and its two directions.
We assign a specific value on a Likert scale ranging from one to five, to paraphrase the sentence as if it were spoken by a person using a language with a certain degree indicated by the score.
This technique can be extended to involve multiple axes and scores (refer to an example result with two axes in \autoref{tab:axes}).
The detailed prompts we used are presented in \autoref{app:para}.

\section{Experiments}
\label{sec5}

\begin{table*}[t]
\caption{\textbf{Accuracy of the generated chart semantics and L1/L2 captions for 48 sample charts (\autoref{fig:level}).} Although 41 out of the 48 sample charts used in our experiment are complex and extra complex, LLMs were able to capture chart semantics and generate L1/L2 captions successfully in general.}
\label{tab:result_accuracy}
\begin{center}
\begin{tabular}{lllc|cc}
\toprule
\multicolumn{4}{c}{\textbf{Metadata}} & \multicolumn{2}{c}{\textbf{Accuracy}}\\
\midrule
& NL Type (\#) & Source & Chart/NL \# & w/ Strict criteria & w/ Lenient criteria\\
\midrule
A. & Chart Semantics (9) & LLM & 48/432 & 89.4\% & 96.9\%\\
B. & L1 Caption (1) & LLM & 48/48 & 76.0\% & 95.8\%\\
C. & L2 Caption (1) & LLM & 48/48 & 76.0\% & 87.5\%\\
\bottomrule
\end{tabular}
\end{center}
\end{table*}

\begin{table*}[t]
\caption{\textbf{Quantitative comparison of benchmarks and LLM-generated utterances and questions.} Two type of metrics were adopted, cross-distribution, which is to compare the two distributions to get the similarity and difference, and within-distribution, which is to compare the diversity within a single distribution. Each NL dataset has come from 4 sources, gold standard or benchmarks, LLM, LLM.P (paraphrased), LLM.P2 (paraphrased with 2 axes). The best metric from all sources are bold, while the best metric in ours (LLM, LLM.P, LLM.P2) are underlines.}
\label{tab:result_diversity}
\begin{center}
\resizebox{\linewidth}{!}{%
\begin{tabular}{lllc|ccc|cccccc}
\toprule
\multicolumn{4}{c}{\textbf{Metadata}} & \multicolumn{3}{c}{\textbf{Cross-Distribution}} & \multicolumn{6}{c}{\textbf{Within-Distribution}}\\
\midrule
& NL Type (\#) & Source & Chart/NL \# & FD ($\downarrow$) & Precision ($\uparrow$) & Recall ($\uparrow$) & RC ($\uparrow$) & Chamfer ($\uparrow$) & MST ($\uparrow$) & Span ($\uparrow$) & Sparsness ($\uparrow$) & Entropy ($\uparrow$)\\ 
\midrule
\multirow{4}{*}{\shortstack{D.}} & \multirow{4}{*}{\shortstack{Utterance (3)}} & Gold & 48/144 & $\cdot$ & $\cdot$ & $\cdot$ & 3.15 & \textbf{0.19} & 47.59 & 3.26 & 2.34 & \textbf{2.60}\\
\cmidrule{3-13}
& & LLM & \multirow{3}{*}{48/144} & 0.58 & 0.81 & 0.31 & 3.31 & \underline{0.19} & 49.81 & 3.41 & 2.48 & \underline{2.54}\\ 
& & LLM.P & & \underline{0.45±0.01} & \underline{0.81±0.01} & 0.66±0.01 & \underline{\textbf{3.65±0.36}} & 0.16±0.01 & \underline{\textbf{54.58±3.63}} & \underline{\textbf{3.52±0.21}} & \underline{\textbf{2.70±0.22}} & 2.09±0.51\\
& & LLM.P2 & & 0.46±0.00 & 0.80±0.02 & \underline{0.67±0.03} & 3.43±0.38 & 0.16±0.01 & 52.91±3.11 & 3.50±0.22 & 2.49±0.20 & 2.27±0.35\\ 
\midrule
\multirow{4}{*}{E.} & \multirow{4}{*}{Question (5)} & Gold & 48/240 & $\cdot$ & $\cdot$ & $\cdot$ & 3.47 & \textbf{0.17} & 70.28 & 3.45 & 2.64 & 2.44\\
\cmidrule{3-13}
& & LLM & \multirow{3}{*}{48/240} & 0.35 & \underline{0.84} & 0.56 & \underline{\textbf{6.20}} & 0.09 & \underline{\textbf{105.16}} & \underline{\textbf{5.92}} & \underline{\textbf{4.40}} & 1.68\\
& & LLM.P & & \underline{0.35±0.00} & 0.76±0.02 & 0.64±0.03 & 4.17±0.20 & 0.13±0.01 & 70.00±3.47 & 4.28±0.30 & 3.13±0.11 & \underline{\textbf{2.51±0.10}}\\
& & LLM.P2 & & 0.36±0.00 & 0.74±0.02 & \underline{0.64±0.03} & 4.29±0.34 & \underline{0.14±0.01} & 77.36±5.91 & 4.44±0.27 & 3.12±0.19 & 2.21±0.34\\
\midrule
\multirow{4}{*}{F.} & \multirow{4}{*}{Utterance (3)} & BM~\cite{srinivasan2021collecting} & 30/804 & $\cdot$ & $\cdot$ & $\cdot$ & 10.42 & \textbf{0.07} & 177.63 & 10.56 & 8.41 & 2.42\\
\cmidrule{3-13}
& & LLM & 30/90 & $\cdot$ & $\cdot$ & $\cdot$ & $\cdot$ & $\cdot$ & $\cdot$ & $\cdot$ & $\cdot$ & -\\
& & LLM.P & 30/804 & 1.11±0.27 & \underline{0.63±0.03} & \underline{0.51±0.05} & \underline{\textbf{12.36±0.18}} & \underline{0.06±0.00} & 209.59±7.07 & 11.74±0.45 & 9.66±0.38 & 2.40±0.06\\
& & LLM.P2 & 30/804 & \underline{0.87±0.56} & 0.58±0.04 & 0.43±0.08 & 12.24±0.18 & 0.06±0.00 & \underline{\textbf{227.70±15.66}} & \underline{\textbf{11.86±0.37}} & \underline{\textbf{9.79±0.19}} & \underline{\textbf{2.45±0.07}}\\
\midrule
\multirow{4}{*}{G.} & \multirow{4}{*}{Question (4)} & BM~\cite{kim2020answering} & 52/629 & $\cdot$ & $\cdot$ & $\cdot$ & 8.66 & \textbf{0.07} & 202.83 & 11.36 & 6.12 & 1.96\\
\cmidrule{3-13}
& & LLM & 52/208 & $\cdot$ & $\cdot$ & $\cdot$ & $\cdot$ & $\cdot$ & $\cdot$ & $\cdot$ & $\cdot$ & -\\
& & LLM.P & 52/619 & \underline{0.33±0.00} & \underline{0.52±0.01} & 0.13±0.01 & \underline{\textbf{11.95±0.27}} & 0.05±0.00 & \underline{\textbf{247.16±11.72}} & 12.46±0.49 & \underline{\textbf{9.07±0.32}} & \underline{\textbf{2.41±0.14}}\\
& & LLM.P2 & 52/629 & 0.33±0.00 & 0.50±0.01 & \underline{0.18±0.01} & 11.61±0.18 & \underline{0.06±0.00} & 222.39±8.21 & \underline{\textbf{12.69±0.24}} & 8.73±0.25 & 2.39±0.06\\
\bottomrule
\end{tabular}
}
\end{center}
\end{table*}

\begin{table}[t]
\caption{\textbf{Types of low-level tasks in questions} Top four ranked in both datasets were identical, while LLM-generated dataset has more questions assigned to these ranks.}
\label{tab:qtype}
\begin{center}
\begin{tabular}{lll}
\toprule
Low-level Analytical Task & Gold \# (\%) & LLM \# (\%)\\ 
\midrule
Retrieve Value & 94 (39.2\%) & 103 (42.9\%)\\
Find Extremum & 35 (14.6\%) & 65 (27.1\%)\\
Correlate & 31 (12.9\%) & 41 (17.1\%)\\
Compute Derived Value & 31 (12.9\%) & 17 (7.1\%)\\
Filter & 23 (9.6\%) & 3 (1.3\%)\\
Find Anomalies & 14 (5.8\%) & 0\\
Characterize Distribution & 5 (2.1\%) & 4 (1.7\%)\\
Sort & 3 (1.3\%) & 0\\
Cluster & 1 (0.4\%) & 0\\
Determine Range & 1 (0.4\%) & 5 (2.1\%)\\
ETC & 2 (0.8\%) & 2 (0.8\%)\\
\midrule
Sum & 240 (100\%) & 240 (100\%)\\
\bottomrule
\end{tabular}
\end{center}
\end{table}

In this section, we introduce quantitative analysis of our generated NL datasets, lexical analysis on generated utterances, and types of low-level tasks in generated questions.

\subsection{Experimental Setup}
Our experiment aims to investigate the effectiveness of our framework in generating diverse NL datasets from Vega-Lite specifications, with a focus on accuracy and diversity. To achieve this, we apply tailored metrics to each NL dataset, taking into account their different characteristics. 
L1/L2 captions are independent of the perception of humans or machines because they focus on conveying objective information \cite{lundgard2021accessible}. Thus, we measured accuracy to determine how precisely each caption level contained relevant information.
We assess the diversity of utterances and questions, as it is important to reflect inclusive language usage among individuals with different background. The results are presented in \autoref{tab:result_accuracy} and \autoref{tab:result_diversity} where each type of NL dataset is classified with capital English letter (A-G).

\subsubsection{Benchmarks}
For utterances and questions, we utilized crowd-sourced NL datasets gathered in prior studies~\cite{srinivasan2021collecting, kim2020answering} (F-BM and G-BM). In case of utterance dataset, we only used the singleton case, so it was 804 sentences instead of 893. However, when it comes to captions, we could not find suitable benchmarks for comparing with different caption levels. Previous research employed bitmap images of charts~\cite{lundgard2021accessible, 2023-vistext}, whereas our approach leverages Vega-Lite specifications. This difference in data format prevented us from making an exact comparison.

\subsubsection{Gold Standard Datasets}
Given that benchmarks mostly focus on simple and medium level complexity with confined diversity, we decided to make a gold standard dataset to test the generalizable performance of our framework over diverse and complex charts.
We referred to previous works~\cite{kong2014extracting, kim2018facilitating} that have demonstrated how to create gold standard datasets. We selected 48 Vega-Lite specifications (\autoref{fig:level}) by stratified sampling, taking into account their complexity level and whether they included interaction or composite views. Subsequently, three visualization experts (first three authors) collaborated to develop three guidelines for generating utterances and questions. These guidelines were crafted by referring to relevant suggestions and guidelines from prior research~\cite{kim2021towards, srinivasan2021collecting, wang2022towards, kim2020answering}.
We began by creating sample utterances and questions for the same chart using the initial drafts, and jointly revised each guideline by reviewing the generated NL datasets. After making consensus about the final guidelines~\autoref{app:gold}, we divided the charts into thirds, with each person tasked with generating NL datasets for their assigned charts. This resulted in 48 utterances (comprising 16 commands, 16 queries, and 16 questions) and 80 questions (including 16 non-visual lookup, 16 non-visual compositional, 16 visual lookup, 16 visual compositional, and 16 open-ended questions) per each expert. After one expert created NL datasets for the assigned charts, the other two individuals conducted verification to find any issues or errors within these generated NL datasets. In cases where issues or errors were detected, all three experts convened to discuss and reach a consensus on how to address them. This collaborative effort resulted in the generation of 144 utterances with three different phrasings and 240 questions categorized into five types (D-Gold and E-Gold).

\subsubsection{LLM-generated Datasets}
To generate our datasets, we used an official API of GPT4\footnote{https://platform.openai.com/docs/models/gpt-4} with the \texttt{gpt-4-0613} model. We set the temperature to 0.0, to solely observe the influence of our paraphrasing technique on diversity. We used different prompt for generating each dataset and paraphrasing the generated NL datasets (see \autoref{app:prompt} and \autoref{app:para}).
Here, we generated all types of chart semantics, captions, utterances, and questions for the 48 sample charts, as well as all types of utterances for 30 charts from the benchmark. This resulted in a total of 432 chart semantics (A-LLM), 48 L1 captions (B-LLM), 48 L2 captions (C-LLM), 144 utterances (D-LLM), and 240 questions (E-LLM) for the 48 sampled charts, and 90 utterances for the 30 benchmark charts (F-LLM). Since the benchmark~\cite{kim2020answering} did not include open-ended questions, we generated only four types of questions. This led to a total of 208 questions for the 52 charts (G-LLM).

We augmented our NL datasets for utterances and questions using the generated NL datasets (*-LLM) and the score-based paraphrasing technique, resulting in augmented paraphrased NL datasets (*-LLM.P and *-LLM.P2). With four language axes and five Likert-scale values (1-5), it is possible to generate 20 different versions (4*5) of paraphrased sentences for each original sentence (i.e., LLM.P). Likewise, in case of two axes, there are six combinations chosen from the four axes. Since there are five Likert-scale options for each axis, this leads to the generation of 150 (6*5*5) different paraphrased sentence versions per original sentence (i.e., LLM.P2). We meticulously generated all possible paraphrases and selected five distinct sets of NL datasets to mitigate any sampling bias. Thus we calculated metrics and their averages and standard deviations across these five sample sets.

When sampling the paraphrased sentences, our goal is to compare the syntactic diversity of different NL datasets while aligning the semantic diversity of the two datasets being compared to ensure a fair comparison. To this end, we adjust the frequency of each chart-NL pair in both datasets. This is necessary because the benchmark data exhibit biases in NL sentence distribution for each chart. For instance, one chart has 30 associated questions, while another chart has only one question. We count the frequency of each chart-NL pair and reflect the same frequency when augmenting the datasets. This became an issue when creating G-LLM.P, since one chart has 30 questions, which exceeds the maximum number of paraphrases possible (limited to 20) through our single-axis paraphrasing method. As a result, our overall number of NL datasets reaches 619.

Last, we included open-ended questions in E-LLM.P and E-LLM.P2, as these questions were available in E-Gold. However, we did not include them in G-LLM.P and G-LLM.P2 in \autoref{tab:result_diversity} to preserve the semantic diversity of the datasets.

\subsubsection{Procedure}
We manually grade chart semantics and L1/L2 captions to compute their accuracy. To enhance the reliability of our scoring, two experts (the first and second authors) independently scored them and calculated the average score. Specifically, the chart semantics include whether they contain composite views, the type of composite view, the number of plots, chart type, mark, transform, encoding, style, and interaction.
We scored whether each of them is correct or not. However, during our evaluation of style, we encountered many cases where multiple width or height values were defined within the Vega-Lite specification. In such cases, we chose to exclude the width and height information from our style evaluation. Moreover, we encountered many cases that were hard to definitively categorize as either correct or incorrect. For instance, situations where nine lines were drawn on the same chart but divided into separate layers, resulting in a count of nine plots instead of one. As a result, we adopted two different scoring approaches, consisting of strict and lenient criteria.
Strict criteria only considers those that were 100\% accurate. For instance, if a stacked bar chart was categorized as a bar chart, it was deemed incorrect. Conversely, with lenient criteria, we adopted a more flexible approach, considering the aforementioned cases as correct. We extended these criteria to the evaluation of L1/L2 captions as well as their formal definitions~\cite{lundgard2021accessible}. As they contain objective information, we applied the same two criteria and reasoning to assess their accuracy.

To assess the quality of utterances and questions in comparison to both the benchmark and the gold standard dataset, we employ two types of statistical metrics: within-distribution and cross-distribution metrics. The within-distribution metrics are designed to calculate the similarity and divergence between a given dataset and another dataset by means of comparison. Examples of such metrics include Frechet distance (FD), precision, and recall. By utilizing these metrics, we can evaluate how closely a given distribution aligns with the benchmark distribution. These metrics have already been applied in the comparison of human-generated and LLM-generated datasets \cite{hamalainen2023evaluating}. To this end, we vectorize the gold standard, benchmarks, and LLM-generated as well as paraphrased datasets, transforming them into sets of vectors for quantitative comparison.

However, we recognize that the aforementioned metrics may not provide a comprehensive measure of the quality of LLM-generated and -paraphrased NL datasets. These metrics mainly focus on the coverage of distribution rather than emphasizing diversity. It is crucial to delve deeper into a single distribution, as duplicate or highly similar data points may be present within it \cite{klein2015high, siangliulue2016ideahound}. To address this, we incorporate cross-distribution metrics \cite{rhys2021directed} that allow us to quantify the diversity within a single distribution. These metrics include remote-clique (average of mean pairwise distances), Chamfer distance (average of minimum pairwise distances), MST dispersion (sum of edge weights of MST), span (Pth percentile distance to centroid), sparseness (mean distance to medoid), and entropy (Shannon-Wiener index for points in a grid partition).

\subsection{Quantitative Results}
We first report the accuracy of chart semantics and L1/L2 captions. Under the strict criteria, the accuracy rates for chart semantics, L1 captions, and L2 captions were 89.4\%, 76.0\%, and 76.0\%, respectively. In detail, accuracy under strict criteria reveals that `chart-type' achieved the lowest accuracy at 75\%, while `mark' and `interaction' showed the highest accuracy at 96.9\%. Under lenient criteria, the accuracy rates for chart semantics, L1 captions, and L2 captions significantly improved to 96.9\%, 95.8\%, and 87.5\%, respectively. Specifically, the lowest accuracy for chart semantics was observed in the `number of plots' (88.5\%), while `mark' and `interaction' maintained the highest accuracy at 100\%. Additionally, the accuracy of chart type substantially improved to 97.9\%. A summary of these results is provided in \autoref{tab:result_accuracy}.

We next report the diversity of utterance and question. In terms of cross-distribution metrics, LLM.P exhibited the highest quality in terms of precision (D), precision and recall (F), and precision and FD (G). In case of datasets containing five question types (E), the metric results were not consistent. Specifically, LLM.P performed the best in FD, LLM was the best for precision, and LLM.P2 achieved the highest recall. When considering within-distribution metrics, LLM-generated and paraphrased datasets demonstrated greater diversity compared to the gold standard and benchmark datasets. On average, higher diversity was observed in 4.75 out of six metrics. For both question and utterance datasets (E, F), paraphrased datasets with two axes demonstrated greater diversity than paraphrased datasets with one axis in four out of six metrics. Conversely, in the other two datasets (D, G), paraphrased datasets with one axis exhibited higher diversity in four out of six metrics. In the utterance dataset (D), paraphrasing increased diversity in four out of six metrics, whereas in the question dataset (E), paraphrasing reduced diversity in four metrics. A summary of the results is presented in \autoref{tab:result_diversity}.

\subsection{Lexical Analysis in Utterances}
To gain a deeper understanding of the syntactic diversity in LLM-generated datasets, we conducted a lexical analysis on three NL datasets (F-BM, F-LLM.P, F-LLM.P2) to investigate the types of words used within each dataset.
Our pre-processing steps encompassed sentence tokenization, converting all text to lowercase, removal of stopwords, and lemmatization.
As evidenced by the quantitative outcome presented in the previous section, the LLM.P exhibited a notable richness in its lexical diversity.
It contained a total of 555 unique words, surpassing the benchmark dataset's count of 349 unique words.
Also, the total word count in the LLM.P, amounting to 7,132 words, exceeded that of the benchmark dataset, which consisted of 4,480 words.
In case of LLM.P2, it demonstrated an even greater number of unique words, totaling 608, surpassing both the benchmark and LLM.P datasets in this regard.
However, the overall word count in LLM.P2 was lower at 6,645 words compared to the LLM.P dataset (7,132 words).

There were some additional patterns in the use of specific words employed within the LLM-generated datasets.
First, the paraphrased dataset introduced a multitude of new action verbs.
For instance, when issuing commands, terms such as construct, fabricate, organize, and arrange were employed to create charts (e.g., `Fabricate a line diagram'). In previous work \cite{srinivasan2021collecting}, there was a tendency among crowd workers to adhere to specific terminology, thus researchers have to be careful when providing instructions for collecting datasets.
Our paraphrasing technique effectively addresses this issue by promoting diverse syntax through the use of various action verbs automatically.
Second, the datasets incorporate words that may be adopted by people of specific groups or domains, but not used often by ordinary people, such as domain-specific jargon (e.g., provenance, bifurcated, barometric, pecuniary).
Last, certain words have been adopted to introduce diverse tones and voices of the speaker.
These encompass terms of a more personal and informal nature, as well as expressions that convey uncertainty and speculation (e.g., maybe, seems, might, quite, sure), as well as words that have been included to enhance conversational aspects (e.g., possibly, would, could).

\subsection{Types of Low-level Tasks in Questions}
Based on a taxonomy~\cite{amar2005low} comprising ten low-level analytical tasks, we conducted an analysis of the question types present in the gold standard and LLM-generated questions (E-Gold and E-LLM).
This analysis aimed to assess the dissimilarities or similarities between these questions.
To this end, we associated each low-level analytical task with individual questions within both datasets.

Both datasets exhibited a congruent pattern, with identical rankings for the top four elements.
The task with the highest frequency in both datasets is retrieve value, which is unsurprising, as it consists of 40\% of lookup questions in the dataset.
Notably, in the LLM-generated dataset, the second most prevalent task is find extremum at 27.1\%.
This percentage closely aligns with Kim et al.'s observation \cite{kim2020answering}, where they reported a similar prevalence of questions related to extrema at 26.7\%.
Furthermore, it is worth highlighting that, akin to their research, there is a clear bias towards certain task types, including retrieve value, find extremum, correlate, and compute derived value \autoref{tab:qtype}.

\section{Application}
\label{sec6}

\subsection{Finetuning LLMs for Data Visualization}
\label{sec6:1}

\begin{table}[t]
\caption{\textbf{The result of finetuning experiment.} The LLMs trained with NL datasets generated by our framework either matched or surpassed the performance of LLMs (C, D, E) compared to when using only the benchmark dataset (A).}
\label{tab:finetuning}
\begin{center}
\begin{tabular}{clccc}
\toprule
& Source & Train \# & Test \# & Accuracy (\#)\\
\midrule
A. & BM~\cite{srinivasan2021collecting}  & 723 & 81 & 76.3\% (61.8)\\
B. & LLM.P & 723 & 81 & 58.8\% (47.6)\\
C. & BM + LLM.P & 723 & 81 & \textbf{76.8\% (62.2)}\\
D. & BM + LLM.P & 1446 & 81 & \textbf{83.2\% (67.4)}\\
E. & BM + LLM.P + LLM.P2 & 2169 & 81 & \textbf{85.4\% (69.2)}\\
\bottomrule
\end{tabular}
\end{center}
\end{table}

We demonstrate that the NL datasets generated by our framework can be used to augment the performance of ML models. It is important to note, however, that the effectiveness of our datasets is contingent upon the availability of a sufficient number of human-generated datasets that exhibit similar distributions to the test datasets. We believe our approach serves as a cost-effective and efficient way to be used in conjunction with the conventional method of crowdsourcing human-generated NL datasets. In essence, our framework's output can be strategically employed as supplementary datasets for finetuning LLMs.

To be specific, to replicate the benchmark dataset's experiment \cite{srinivasan2021collecting}, we performed an experiment to classify ten chart types (e.g., colored scatterplots, stacked \& grouped bar charts, multiseries line charts, etc.) using utterances. This classification task is important as it can be further used for building visualization systems like chart type recommendation. We prepared five datasets for finetuning: A. the benchmark dataset (723 utterances), B. the utterances generated and paraphrased with one axis by our framework (723 utterances), C. half of A and half of B (362 from A + 361 from B), D. A and B (723 from A + 723 from B), E. D as well as the utterances generated and paraphrased with two axes by our framework (1446 from D + 723 additional utterances). Only 90\% of the benchmark dataset is used for finetuning and the rest 10\% were used for the test. Similarly, we used only 90\% of our datasets to maintain an equal number of utterances as in the benchmark dataset. Following common ML practices, we selected OpenAI's \texttt{babbage-002} model for training smaller models on downstream tasks, setting hyperparameters to default configurations (i.e., number of epochs as 3, learning rate multiplier as 2, and the batch size were 1, 1, 1, 2, 4 for each case). Each experiment was repeated five times to calculate the average accuracy to mitigate the stochastic behavior of LLMs.

As denoted in \autoref{tab:finetuning}, we observed an increase in performance when using LLM-generated NL datasets alongside the benchmark dataset for finetuning the models. Using only the benchmark datasets resulted in an accuracy of 76.3\% (61.8 accurate prediction on average out of 81, \autoref{tab:finetuning}-A). When we combined the benchmark datasets with our dataset, the accuracy slightly improved to 76.8\% (62.2 out of 81, see \autoref{tab:finetuning}-C), indicating that the addition of a non-human-generated datasets did not negatively impact accuracy. Moreover, the performance increased to 83.2\% when we leveraged additional NL datasets generated by LLMs (67.4 out of 81, \autoref{tab:finetuning}-D). The accuracy was the highest when we used more NL datasets paraphrased with two language axes by our framework, which is 85.4\% (69.2 out of 81, \autoref{tab:finetuning}-E). Last, using only LLM-generated NL datasets showed decreased accuracy, which is 58.8\% (47.6 out of 81, \autoref{tab:finetuning}-B).

The results suggest that using the NL datasets, generated and paraphrased by our framework, can enhance the performance of ML models in downstream tasks. We believe a key factor in this improved performance is the increased syntactic diversity of the generated utterances, which also accurately mimic semantic characteristics. Our results align with a previous finding that utilizing AI-generated datasets can become a more cost-effective strategy for training scalable ML models with significantly fewer human labels \cite{bai2022constitutional}. This suggests that the synergistic use of both human efforts and our automated framework can substantially enhance the quality of training data and the performance of the models.

\subsection{Leveraging Fully-automatic and Mixed-initiative Modes in VL2NL}

\begin{figure*}[t]
  \centering
  \includegraphics[width=\linewidth]{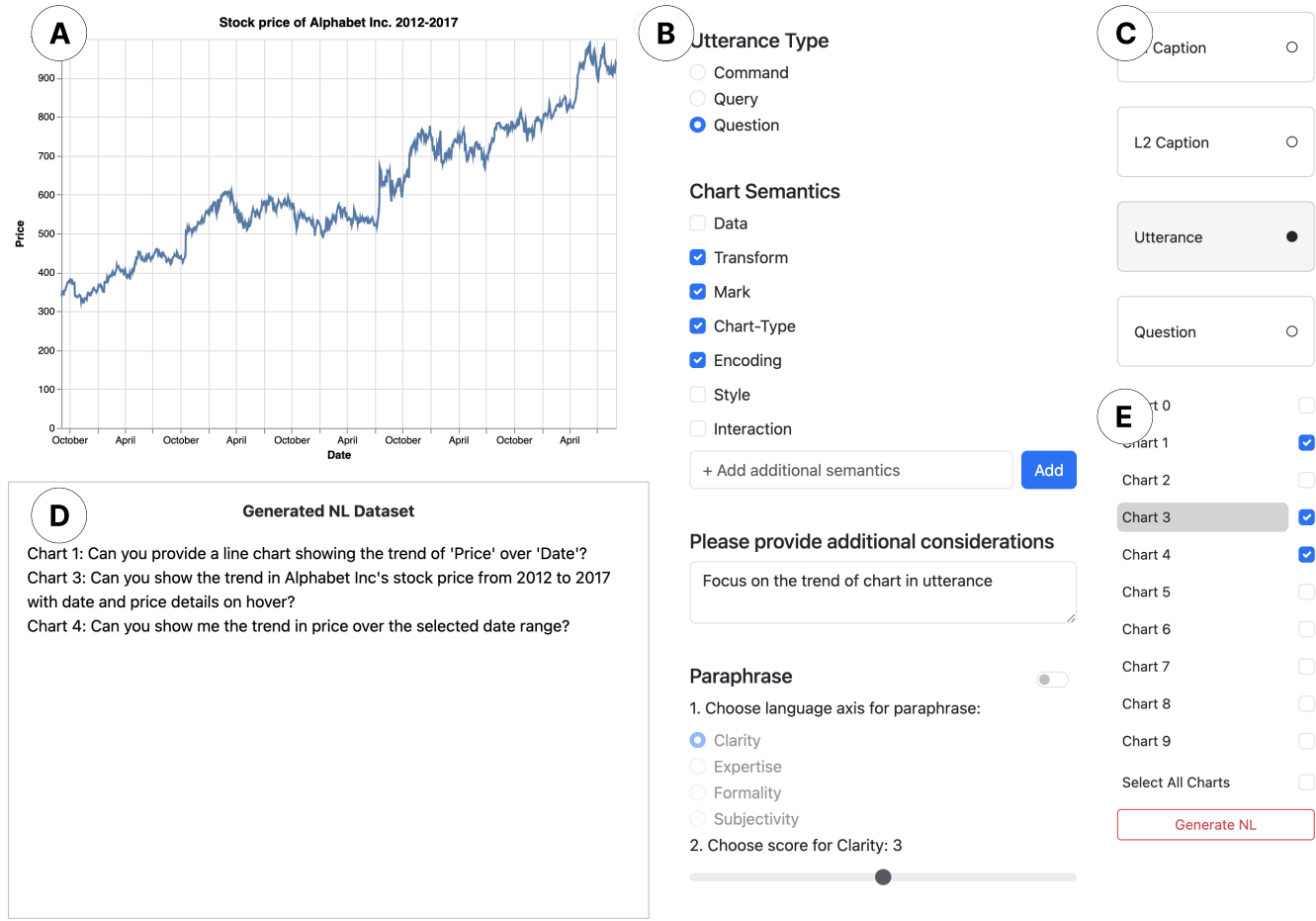}
  \caption{\textbf{A system with two modes (fully-automatic and mixed-initiative) to generate NL datasets using VL2NL.} The mixed-initiative mode encompasses several features. First, users can select the types of NL datasets they want to generate (C). They can inspect each chart (A) and subsequently choose the specific ones they wish to use for generating NL datasets (E). Users can change or provide information that the system utilizes (B). Once these are completed, the system returns the generated NL datasets (D). In contrast, the fully-automatic mode does not include (B). As a result, dataset generation in this mode strictly follows the scaffolding defined by researchers, along with key questions and answers generated by LLMs.}
  \label{fig:system}
\end{figure*}

To further explore how visualization researchers can use our framework, we performed a case study with two experts: E1, a professor, and E2, a postdoctoral researcher. Both have earned their Ph.D. in visualization and have conducted research for 10 and 7 years, respectively.

For the case study, we implemented a system with two modes: fully-automatic and mixed-initiative (\autoref{fig:system}). In the fully-automatic mode, the scaffolding is set by us and key questions are generated automatically by the LLMs. Therefore, users had no control, but could click the button to generate NL datasets for their chosen charts. In the mixed-initiative mode, users can select which scaffolding to consider and provide additional information as answers to key questions. They can actively contribute by specifying directions to steer its focus accordingly. For example, in case of L1 caption, they can choose which chart semantics to consider or add more when generating it. Similarly, for question, users can make a high-level decision themselves, specifying where or what to focus on when analyzing the charts.

To clarify our study protocol, we first provided the experts with an overview of our framework's concept. Following this, they were given a task to create NL utterances for 10 line charts, which depicted stock prices of various technology companies \cite{xu2018stock}. This was conducted using two modes: fully-automatic and mixed-initiative. We emphasized to the participants that the utterances they generated would be instrumental in training an ML model to translate these utterances back into the corresponding line charts. We also highlighted the significance of utterance diversity in enhancing the performance of ML models, based on our discussion in (\autoref{sec6:1}). Finally, the experts provided feedback on their anticipated use of both modes for generating utterances. Each expert spent approximately 45 minutes for the study.

Both experts agreed on using both modes to generate utterances more effectively for training ML models. Specifically, E1 suggested the following scenario: initially, researchers generate a large number of utterances automatically to observe their distribution. Next, they identify areas lacking in diversity, which then become the focus for generating additional utterances subsequently. By repeatedly testing and generating utterances in these sparse areas, particularly using the mixed-initiative mode, they can achieve a more diverse and evenly distributed utterances. This process, iterated over multiple times, could improve the performance of the ML models. Similarly, E2 also advocated starting with the fully-automatic mode before using the mixed-initiative mode. E2 said this approach allows experts to better understand the model's behavior and the nature of the utterances it generates. This step is crucial to avoid `option paralysis,' a state of cognitive overload that may occur when faced with a lot of choices without a clear strategy for improvement. With a deeper understanding of the model's behavior, they can proceed more effectively.

\section{Discussion}
\label{sec7}

\begin{figure*}[t]
  \centering
  \includegraphics[width=\linewidth]{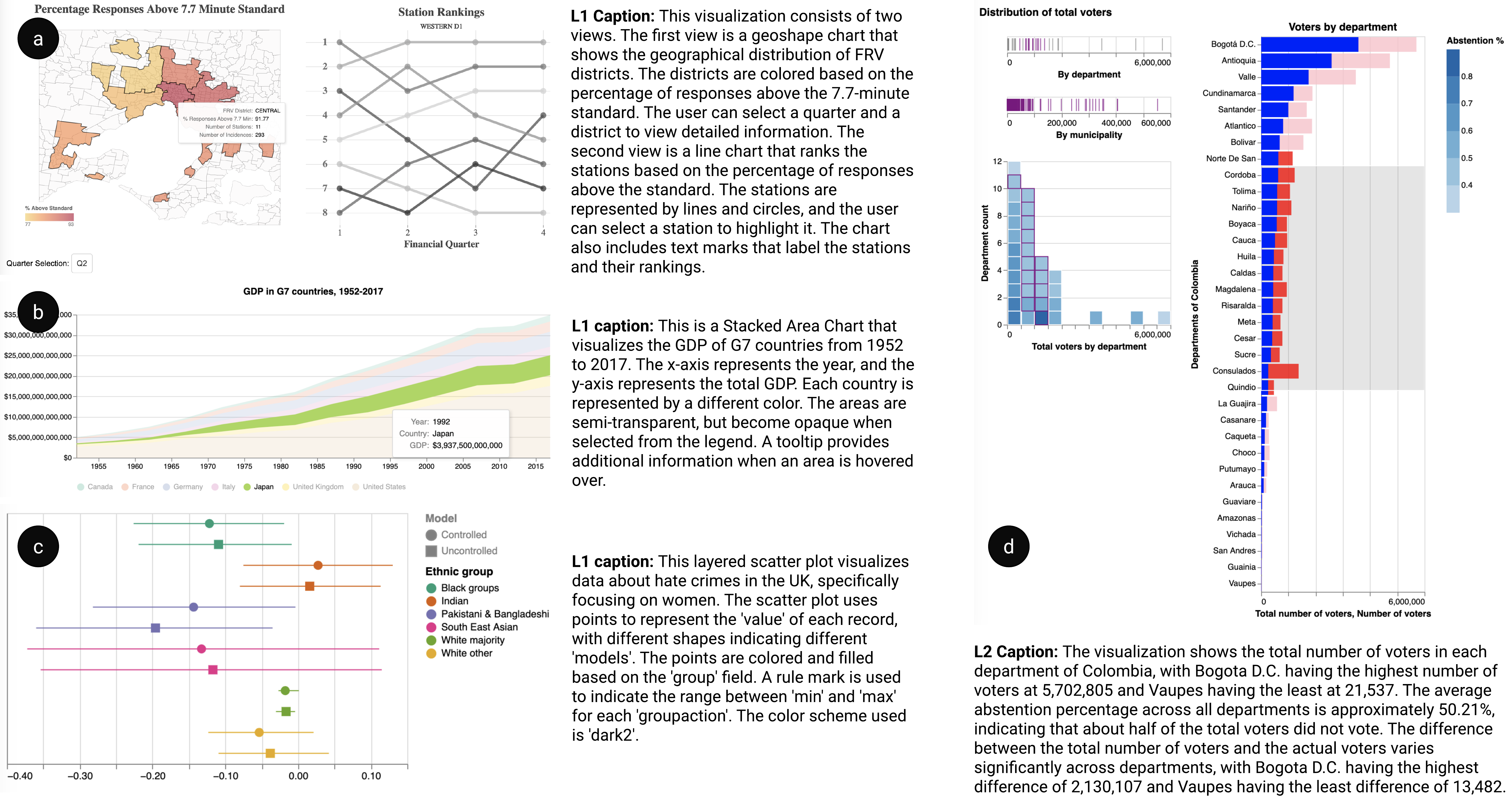}
  \caption{\textbf{Four examples of generated L1/L2 captions with corresponding charts.} We found that VL2NL can successfully generate captions even on complex charts with varying interactions and multiple views.}
  \label{fig:l1l2}
\end{figure*}



\subsection{Strengths and Weaknesses of VL2NL}

\subsubsection{VL2NL Can Guide Itself via Key Questions}
We observed several interesting key questions discovered by LLMs that play a vital role in guiding themselves.
They were formulated through a meticulous analysis of chart contents.
Various decision-making processes were identified, spanning diverse domains such as financial decision-making (e.g., assessing whether to invest in a company's stock), public policy planning (e.g., formulating policies based on employment trends across different age groups and countries), and location-based business strategies (e.g., selecting optimal sites for a new shoe factory relative to the distribution of existing facilities).
These key questions served as the foundation for eliciting subsequent conclusions, retrieving specific values, and deciding mathematical operations for generating interesting questions.

\subsubsection{VL2NL Works Robustly on Different Chart Complexity}
Considering that the 48 sampled charts mostly belong to the categories of complex and extra-complex charts, our observations indicate that the reported accuracy (\autoref{tab:result_accuracy}) pertaining to chart semantics and L1/L2 captions does not exhibit a dependence on the complexity levels of the Vega-Lite specifications.
This finding underscores the robustness of the system.
For instance, it successfully generated an accurate L1 caption for a chart comprising two views interconnected through selection interactions (see \autoref{fig:l1l2}-a).
Similarly, it effectively generated an L2 caption for a chart containing multiple plots, allowing the selection of a data range in the main bar plot to trigger the highlighting of related data points in other plots (see \autoref{fig:l1l2}-d).

\subsubsection{VL2NL Depends Highly on Vega-Lite Specifications}
We observed that the framework is highly dependent on the Vega-Lite specifications in generating NL datasets.
In many cases, this dependency is advantageous as it enables a focus on intricate functionalities such as interactions.
For a particular chart, determining the presence of interactions was challenging because the selection interaction was indicated solely by the color label of the chart.
Nevertheless, the framework successfully captured this (see \autoref{fig:l1l2}-b).
Similarly, charts lacking titles or descriptions can pose a challenge in comprehending the content of charts.
However, it appears that the framework was able to extract additional information, even utilizing the URL of the data included in the specification (e.g., \texttt{how-did-levels-of-uk-hate-crime-change-during-and-after-covid-19/data/f5.csv}), enabling the generation of informative and coherent captions (see \autoref{fig:l1l2}-c).

However, we have identified certain cases where relying solely on Vega-Lite specifications proves disadvantageous.
First, in some instances, the generated NL datasets include information that was not visually represented in the chart but was present in the Vega-Lite specifications.
For instance, additional categories or values that exceed the specified axis range were presented in the NL dataset.
Second, if certain information is not explicitly stated within the Vega-Lite specification, it cannot be incorporated into the NL dataset.
For example, when generating trellis plots, the number of plots is determined using the unique count of elements.
However, since the number is not explicitly provided in the specification, our framework is unable to predict the exact plot number accurately.
Last, any errors present in the Vega-Lite specifications are faithfully represented in the generated NL datasets.
For instance, a specification contained a typo that divided facets into 3 parts but was mistakenly denoted as 4, our framework predicted the number of plots as 4 instead of the correct 3.
Similarly, the Vega-Lite specification included code that is non-functional, which is reflected in the generated captions, resulting in inaccuracies.

\subsubsection{VL2NL Predicts Only Common Chart Types}
Our categorization relied on the chart type taxonomy presented by Borkin et al.~\cite{borkin2013makes}, which led to different categorizations even when the same mark was used.
For instance, although the same point mark was employed, it could be interpreted as either a distribution chart (e.g., dot array) or a scatter plot.
Furthermore, we conducted an analysis of the charts by considering their detailed sub-types rather than grouping them into larger categories.
For instance, a chart featuring a stacked area chart was considered incorrect according to strict criteria if it was predicted as an area chart.
However, we observed that in most cases, the LLM framework tended to assign the charts to the most prevalent and common chart types such as scatterplots, area charts, and bar charts, rather than classifying them as distribution charts, stacked area charts, or stacked bar charts (\autoref{fig:l1l2}-c).
With this reasons, the chart type exhibited the highest accuracy gap between strict and lenient criteria.



\subsection{Limitations and Future Work}

\subsubsection{Enriching Capabilities of VL2NL through External Resources}
While Vega-Lite specifications serve as powerful inputs for generating various types of NL datasets, it is inherently challenging to extract information that does not exist within these specifications. Although our framework can operate in both fully-automatic and mixed-initiative manner, it does not rely on external resources. This limitation can potentially impact the performance of NL generation, as it aligns with observations in guided discovery, where insufficient prior knowledge can hinder learners from formulating hypotheses, interpreting data effectively, and engaging in systematic experimentation \cite{de1998scientific}. To enhance the capabilities of VL2NL, we suggest accessing external information to guide the process during NL dataset generation. For instance, generating L3/L4 captions often necessitates access to common or domain-specific knowledge \cite{lundgard2021accessible}. In this regard, employing tools like ReAct~\cite{yao2023react} becomes advantageous, as it facilitates reasoning to assist the model in deducing, tracking, and updating action plans while also handling exceptions. This enables us to proactively retrieve information from the web when required.

\subsubsection{Augmenting Vega-Lite Specifications}
While we have presented the largest amount of Vega-Lite specifications and acknowledge their ability as input for generating diverse NL datasets, it is noteworthy that the quantity of Vega-Lite specifications is significantly smaller compared to bitmap images. This is mainly because collecting Vega-Lite specifications is more challenging when compared to other formats. This limitation hinders the effective training or fine-tuning of machine learning models to achieve robust performance. Consequently, we posit the need for methods to augment Vega-Lite specifications. Various augmentation techniques have been introduced and adopted for bitmap images of charts to increase both their quantity \cite{jung2017chartsense} and diversity \cite{zhao2020chartseer}. However, to the best of our knowledge, we have not found any pertinent research that addresses the augmentation of Vega-Lite specifications. As part of our future work, we aim to tackle this gap by developing a reverse engineering technique~\cite{poco2017reverse} specifically designed for Vega-Lite specifications.

\subsubsection{Covering Additional NL datasets}
Our framework exhibits potential for generalization across multiple NL datasets. However, we recognize that it covers only limited number of types, which we aim to expand in our future research. Specifically, we plan to create conversational NL datasets to facilitate interactive communication with NLIs, given the growing significance of conversational agents. We believe a dataset based on deeper analysis of users' conversational characteristics will be immensely beneficial for researchers. We also plan to address reference datasets linking charts with text to help make interactive documents. We believe this will make the connection between them clearer, and the reading experience more enjoyable and engaging.

\section{Conclusion}
We introduce VL2NL designed to generate diverse NL datasets aimed at enhancing NLIs for data visualization research. Our framework takes a Vega-Lite specification as input and employs guided discovery to accurately generate various NL datasets, including captions, utterances, and questions. We also propose a score-based paraphrasing approach to enhance the syntactic diversity of the generated NL datasets. We also present a dataset comprising 1,981 Vega-Lite specifications. This dataset surpasses the baselines in terms of complexity and diversity. Our experimental results substantiate that the framework excels in accurately generating both L1 and L2 captions, while achieving higher diversity in the generation of utterances and questions compared to the baselines. Last, we introduce real-world scenarios of using LLM-generated NL datasets and our framework. We hope our framework and chart collection can advance research in developing NLIs for data visualization.

\begin{acks}
This work was supported by the National Research Foundation of Korea (NRF) grant funded by the Korea government (MSIT) (No. 2023R1A2C200520911). This work was also supported by Institute of Information \& communications Technology Planning \& Evaluation (IITP) grant funded by the Korea government (MSIT) (No.2019-0-00075, Artificial Intelligence Graduate School Program (KAIST)).
\end{acks}

\bibliographystyle{ACM-Reference-Format}
\bibliography{10.reference}

\appendix

\section{Prompts for NL Generation}
\label{app:prompt}

In the prompt, certain variables are enclosed within curly brackets. We colored them blue for easy recognition. Here, we provide a detailed explanation of each variable and specify its usage in different NL generation prompts:

\begin{itemize}
  \item \texttt{vl} [all]: Minified Vega-Lite specification;
  \item \texttt{ftt\_str} [L1/L2 caption, utterance]: Information about fields, titles, types, and values;
  \item \texttt{prompt} [L2 caption]: Questions derived from the guided discovery process;
  \item \texttt{info} [L2 caption]: Answers for the questions computed through LangChain library~\cite{Chase_LangChain_2022};
  \item \texttt{info\_first\_concat} [utterance]: A list of primary instructions by analyzing the chart semantics.
\end{itemize}

\subsection{L1 Caption}

\begin{lstlisting}
{vl}

Let's generate a level 1 NL description step by step.

Step 1. Determine if the visualization contains composite views, such as layered plots, trellis plots, or other types of multiple view displays, and provide a count of the number of plots if any are present.
Step 2. Analyze the semantics of each chart individually, including [Data], [Transform], [Mark], [Chart-Type], [Encoding], [Style], and [Interaction]. Refer to this:
{ftt_str}
Step 3. Generate a level 1 NL description using the semantics. It contains elemental and encoded properties of the visualization (i.e., the visual components that comprise a graphical representation's design and construction).

##
Step 1. Composite Views:
- True/False:
- (If True) Type: (layered, trellis, multiple views)
- Number of plots:
Step 2. Chart Semantics:
- Data:
- Field (Value):
- Transform:
- Mark:
- Chart-Type:
- Encoding:
- Style:
- Interaction (e.g., tooltip):
Step 3. Level 1 NL Description:
\end{lstlisting}

\subsection{L2 Caption}

\begin{lstlisting}
{vl}

Let's generate question(s) step by step.

Step 1. What is the most prominent and meaningful feature in the given chart?
Step 2. What is the mathematical operation(s) (e.g., max, min, sum, difference, and average) required to describe the feature?
Step 3. Generate question(s) using the mathematical operation(s) required to describe the feature. If there are multiple questions, separate them with semicolon(;).

##
Step 1. Features:
Step 2. Operations:
Step 3. Questions:
\end{lstlisting}

\begin{lstlisting}
Refer to this: {ftt_str}
Do not draw any charts to answer the question.

Question: {prompt}
\end{lstlisting}

\begin{lstlisting}
Information: {info}

{ftt_str}

Generate a concise level 2 NL description of a visualization, with 1 or 2 sentences. It contains statistical and relational properties of the visualization (e.g., descriptive statistics, extrema, outliers, correlations, point-wise comparisons).

##
Level 2 NL Description:
\end{lstlisting}

\subsection{Utterance}

\begin{lstlisting}
{vl}

Step 1. Determine if the visualization contains composite views, such as layered plots, trellis plots, or other types of multiple view displays, and provide a count of the number of plots if any are present.
Step 2. Provide a list of instructions to create the chart using natural language.
- Write instructions for each view and separate with <%>
- Separate each instruction by a semicolon (;)
- Divide each instruction to contain only one specific action
- Use the following chart semantics to specify instructions: [Data], [Chart-Type], [Mark], [Encoding], [Transform], [Style], [Interaction]
Step 3. Given the information about the fields and their synonyms, please replace the field names with their corresponding synonyms.
{ftt_str}

##
Step 1. Composite Views:
- True/False:
- (If True) Type: (layered, trellis, multiple views)
- Number of plots:
Step 2. Instructions:
[View #]; [<Chart Semantic>]: <Instruction>; [<Chart Semantic>]: <Instruction>; ... <%>
Step 3. Instructions:
[View #]; [<Chart Semantic>]: <Instruction>; [<Chart Semantic>]: <Instruction>; ... <%>
\end{lstlisting}

\begin{lstlisting}
{inst_first_concat}
The above are instructions to generate a chart. Let's generate combined instructions ([Command], [Query], [Question]) for each view step by step.

Step 1. Identify the primary information in each view.
Step 2. Identify the secondary information in each view.
Step 3. Generate a [Command] for each view using only the primary info. Please follow these rules:
- Use imperative voice
- Write in a single sentence
- Use only the primary info
- Make it concise and simple
Step 4. Generate a [Query] for each view using only the primary info. Please follow these rules:
- Refrain from using verbs and articles (e.g., a, the)
- Use only variables, fields, attributes, mathematical formulas (e.g., sum, avg, mix, max, count, order), abbreviations (e.g., vs), and prepositions (e.g., of, by, for, with, over, from, to)
- Write in a single sentence
- Use only the primary info
- Make it concise and simple
Step 5. Generate a [Question] for each view using only the primary info. Please follow these rules:
- Ask an inquiry as a question
- Write in a single sentence
- Use only the primary info
- Make it concise and simple

##
View #<Number>:
Step 1. Primary Information:
Step 2. Secondary Information:
Step 3. Command:
Step 4. Query:
Step 5. Question:
\end{lstlisting}

\subsection{Question}

\begin{lstlisting}
{vl}

Let's generate a lookup question, a compositional question, and an open-ended question for a given Vega-Lite spec step by step. The lookup question requires a single value retrieval. The compositional question requires multiple operations.

Step 1. What higher-level decision can be made by analyzing this chart?
Step 2. What is a possible conclusion that can be reached from this decision?
Step 3. What specific value can be retrieved to reach this conclusion?
Step 4. Generate a lookup question using this value, without including any visual attributes such as color, length, size, or position.
Step 5. What visual attributes are required to paraphrase this question?
Step 6. Paraphrase the generated question using the chart's visual attributes.
Step 7. What are the mathematical operations (e.g., max, min, sum, difference, and average) to reach the conclusion in Step 2?
Step 8. Generate a compositional question using these operations, without including any visual attributes such as color, length, size, or position.
Step 9. What visual attributes are required to paraphrase this question?
Step 10. Paraphrase the generated question using the chart's visual attributes.
Step 11. Generate an open-ended question to reach the conclusion in Step 2.

##
Step 1. Decision:
Step 2. Conclusion: 
Step 3. Specific Value:
Step 4. Lookup Question: 
Step 5. Visual Attributes: 
Step 6. Paraphrased Question: 
Step 7. Operations: 
Step 8. Compositional Question: 
Step 9. Visual Attributes: 
Step 10. Paraphrased Question: 
Step 11. Open-ended Question: 
\end{lstlisting}

\section{Prompt for Automatic Qualitative Coding}
\label{app:aqc}

When extracting codes, we omitted the words `language' and `use of' since they were frequently added to the code. We believe that these additions do not contribute any additional meaning to the thematic analysis.

\begin{lstlisting}
Let's perform a thematic analysis in the field of human-computer interaction. Generate characteristics of languages leveraged in the given sentence. The total number is five and each of them is separated by semicolons. Do not add numbering or any explanations.

Sentence: {sent}

##
; ; ; ;
\end{lstlisting}

\section{Prompts for Score-based Paraphrasing}
\label{app:para}

We explain the variables used in our prompts:

\begin{itemize}
  \item \texttt{Example Sentence}: A sentence we want to paraphrase;
  \item \texttt{Axis}: An explanation about each of the defined language axes;
  \item \texttt{Direction}: A set of two opposite directions of the given language axis;
  \item \texttt{Score}: A specific value on a Likert-scale ranging from one to five assigned to each of the language axis.
\end{itemize}

\subsection{Paraphrasing with one axis}

\begin{lstlisting}
{Axis}

Score of 1 indicates a higher tendency to use {Direction-1} language and a Score of 5 indicates a higher tendency to use {Direction-2} language. Rewrite the following sentence as if it were spoken by a person with a given score for language usage.

Sentence: {Example Sentence}
Score: {Score}
\end{lstlisting}

\subsection{Paraphrasing with two axes}

\begin{lstlisting}
{Axis-1}
{Axis-2}

Score-A of 1 indicates a higher tendency to use {Direction-1-1} language and a Score-A of 5 indicates a higher tendency to use {Direction-1-2} language.
Score-B of 1 indicates a higher tendency to use {Direction-2-1} language and a Score-B of 5 indicates a higher tendency to use {Direction-2-2} language.
Rewrite the following sentence as if it were spoken by a person with a given score for language usage.

Sentence: {Example Sentence}
Score-A: {Score-A}, Score-B: {Score-B}
\end{lstlisting}

\section{Gold Reference Guidelines}
\label{app:gold}


\subsection{Utterance}
\begin{itemize}
  \item Imagine writing utterances to display a visualization using a system like Excel, Tableau, or Microsoft Power BI;
  \item Refer to both the dataset and the chart to better understand the context in which the data has been used for the visualization and formulate more naturalistic utterances.
  \item Avoid referring to specific instructions to prevent acclimatization to the words or phrases in the instruction \cite{srinivasan2021collecting};
  \item Write utterances as singletons, which are basic types of utterances, but can be more than one sentence if necessary due to complexity, forming a sequential utterance that provides all necessary information;
  \item Write utterance for each view. If the chart is has layered plots, and they have different chart types, write utterance with according to the number of different chart types;
  \item Focus on primary information such as chart type and encoding rather than secondary information such as style and interaction \cite{wang2022towards}.
\end{itemize}

\subsection{Question}
\begin{itemize}
  \item Ask one question in one complete sentence;
  \item Keep questions clear and concise, avoiding overly broad or vague questions by focusing on specific aspects of the chart;
  \item Formulate questions that can elicit useful insights from the visualization to facilitate visual data analysis and decision-making \cite{kim2020answering}.
\end{itemize}

\end{document}